\documentclass[a4paper,11pt]{article}
\pdfoutput=1 

\usepackage{jheppub} 
\usepackage{siunitx}                     
\usepackage{cleveref}                     

\usepackage[T1]{fontenc} 
\usepackage{import}
\usepackage{physics}
\usepackage{siunitx}
\usepackage{caption}
\usepackage{subcaption}
\usepackage{collcell}

\newcommand{\gluon}{g}
\newcommand{\quark}{q}
\newcommand{\squark}{{\tilde q}}
\newcommand{\upsquark}{{\tilde u}}

\newcommand{\gluino}{{\tilde g}}
\newcommand{\gaugino}{{\tilde \chi}}

\title{Soft gluon resummation for associated squark-electroweakino production at the LHC}

\author[a]{Juri Fiaschi}
\author[b]{\!\!, Benjamin Fuks}
\author[c]{\!\!, Michael Klasen}
\author[c]{and Alexander Neuwirth}

\affiliation[a]{Department of Mathematical Sciences, University of Liverpool, Liverpool L69 3BX, United Kingdom}
\affiliation[b]{Laboratoire de Physique Th\'eorique et Hautes \'Energies (LPTHE), UMR 7589, Sorbonne Universit\'e et CNRS, 4 place Jussieu, 75252 Paris Cedex 05, France}
\affiliation[c]{Institut  für  Theoretische  Physik,  Westfälische  Wilhelms-Universität  Münster,  Wilhelm-Klemm-Straße 9, 48149 Münster, Germany}

\emailAdd{fiaschi@liverpool.ac.uk}
\emailAdd{fuks@lpthe.jussieu.fr}
\emailAdd{michael.klasen@uni-muenster.de}
\emailAdd{alexander.neuwirth@uni-muenster.de}

\abstract{
We perform a threshold resummation calculation for the associated production of squarks and electroweakinos at the LHC to the next-to-leading logarithmic (NLL) accuracy. Analytical results for the process-dependent soft anomalous dimension and the hard matching coefficient are presented. The resummed results are matched to fixed-order predictions at next-to-leading order (NLO) in QCD, which are generalised to scenarios with non-universal squark masses and mixings. Numerically, the NLL contributions increase the total NLO cross section by $2\%$ to $6\%$ for squark masses ranging from 1 TeV to 3 TeV, respectively, and they reduce the dependence of the predictions on the factorisation and renormalisation scales from typically $\pm10\%$ to below $\pm5\%$. Our NLO and NLO+NLL calculations have been implemented in the publicly available program \textsc{Resummino}.
}

\preprint{MS-TP-22-05, LTH 1299}


\begin{document} 
\maketitle
\flushbottom

\section{Introduction}
\label{sec:intro}

In the search for physics beyond the Standard Model (SM), supersymmetry (SUSY) offers a very promising avenue \cite{Wess:1974tw,Salam:1974yz,Ferrara:1974pu}. By relating bosons and fermions, it stabilises the Higgs boson mass with respect to quantum corrections, allows for the unification of the gauge couplings \cite{Sakai:1981gr,Chamseddine:1982jx,Ellis:1990wk} and provides a dark matter candidate \cite{Ellis:1983ew,Jungman:1995df}, thus solving several long-standing problems of the SM simultaneously. Within the Minimal Supersymmetric Standard Model (MSSM)~\cite{Fayet:1976cr,Inoue:1982pi,Nilles:1983ge,Haber:1984rc}, each degree of freedom of the SM is associated with a superpartner, that differs only by half a unit in spin. The scalar partners of the left- and right-handed quarks mix, as do the higgsino, bino and wino interaction eigenstates, which after electroweak symmetry breaking form neutral and charged so-called electroweakinos (neutralinos and charginos). Since these particles have not yet been discovered, it must be assumed that supersymmetry is broken \cite{Witten:1981nf,Girardello:1981wz}, which increases the masses of the supersymmetric particles with respect to those of their SM partners. With the assumption of conserved $R$-parity \cite{Farrar:1978xj}, a neutral lightest supersymmetric particle (LSP) then has all the right properties to be one of the most promising dark matter candidates, {\it i.e.}~a weakly interacting massive particle (WIMP).

Both the upcoming Run 3 of the Large Hadron Collider (LHC) and its planned extension to high luminosity (HL-LHC) will provide access to very massive new particles~\cite{hllhc, CidVidal:2018eel, Gianotti:2002xx}. In supersymmetry, the generally dominant production processes are those involving the strong interaction and thus concern the pair production of squarks and gluinos. As a consequence of the current and expected bounds on these states, they might be too massive to be pair-produced at the LHC. This restriction can be lifted by considering the single production of a squark or a gluino in association with a (typically lighter) electroweakino. These processes then also have the advantage of providing insights not only on the supersymmetric masses, but also on the supersymmetric interactions~\cite{1108.1250,1907.04898,2110.04211}.

Only precise theoretical predictions allow for a reliable comparison between theory and experiment. Already by going from leading order (LO) to next-to-leading order (NLO) in perturbative QCD, the theoretical uncertainty originating from the arbitrary choice of factorisation and renormalisation scales is reduced. However, with the possibility of light SUSY particles being excluded by direct searches at the LHC, the current mass limits imply that in any SUSY production process the kinematic configuration approaches the production threshold. This results in large threshold logarithms ruining the convergence of the perturbative series, so that they must be resummed. This resummation procedure has been known for quite some time at the leading logarithmic (LL) and next-to-leading-logarithmic (NLL) accuracy and in some cases beyond and has been found to generally further reduce the theoretical uncertainty inherent in the perturbative calculation~\cite{Sterman:1986aj,Catani1989, Catani1991,Kidonakis:1997gm,hep-ph/9801268}.

In the last decade the precise investigation of slepton pair \cite{Bozzi:2004qq,Bozzi:2006fw,Bozzi:2007qr,Bozzi:2007tea,Fuks:2013lya,Fiaschi:2018xdm,Fiaschi2020}, electroweakino pair~\cite{Debove:2008nr,Debove:2009ia,Debove:2010kf,Debove:2011xj,Fuks:2012qx,Fuks:2013vua,1805.11322,Fiaschi:2020udf} and electroweakino-gluino \cite{1604.01023} production beyond LO \cite{Barger:1983wc,Dawson:1983fw} and NLO \cite{PhysRevLett.83.3780,Berger:1999mc,Berger:2000iu,Spira:2002rd} was accomplished. Similarly, improved predictions in the strong sector exist for squark and gluino pair production~\cite{Beenakker:1994an,Beenakker:1995fp,Beenakker:1996ch,Beenakker:1997ut,Kulesza:2008jb,Beenakker:2011sf,1601.02954,Beenakker:2009ha,Beenakker:2013mva,Beenakker:2014sma,1510.00375}. In this work we focus on the production of a squark and an electroweakino at hadron colliders. Since this process involves both weak [${\cal O}(\alpha_\text{EM})$] and strong [${\cal O}(\alpha_s)$] interactions at leading order, the resulting cross section is of intermediate size. The simplest process in this category involves the production of a first- or second-generation squark together with a lightest neutralino. This process manifests itself through a hard jet originating from the squark decay and missing transverse energy from the two neutralinos leaving the detector invisibly, one of them being a decay product of the squark and the other one being directly produced in the hard process. Such a monojet signal is particularly well-studied in the context of dark matter production at colliders~\cite{Feng:2005gj,Bai:2010hh}. In recent analyses by the CMS collaboration, squark masses below \SI{1.6}{TeV} were excluded in four mass-degenerate squark flavour models, assuming production with a light neutralino $\gaugino^0_1$. This limit is reduced to \SI{1.1}{TeV} for a single kinematically reachable squark \cite{CMS:2017abv, 1908.04722, CMS:2021cox}.  Similarly, the ATLAS collaboration gives limits of \SI{1.4}{TeV} and \SI{1.0}{TeV}~\cite{2010.14293, 2101.01629}.

In this paper, we present a threshold resummation calculation for the associated production of squarks and electroweakinos at the NLO+NLL accuracy. The structure of this work is as follows: In \cref{sec:soft}, we compute the production cross section at leading and next-to-leading order and give a brief summary of the ingredients required for threshold resummation. The numerical validation of our NLO calculation and our new results up to NLO+NLL accuracy are given in \cref{sec:num} for various benchmark scenarios. We summarise our work and conclude in \cref{sec:concl}.

\section{Soft gluon resummation}
\label{sec:soft}

We begin this work with a derivation of LO and NLO expressions for the associated production of a squark and an electroweakino at hadron colliders in \cref{sec:lo_and_nlo}. Then, \cref{sec:ref} explains refactorisation and resummation up to NLL accuracy and includes a computation of the soft anomalous dimension associated with the process considered. In \cref{sec:hard} we present the NLO hard matching coefficient, which is then used together with the soft anomalous dimension in \cref{sec:match} to consistently combine fixed-order and resummed predictions at the NLO+NLL accuracy.

\subsection{Production of squarks and electroweakinos at leading and next-to-leading order}
\label{sec:lo_and_nlo}

To calculate the total hadronic cross section $\sigma_{AB}$ for the process considered, we convolve the partonic cross section ${\rm d}\sigma_{ab}$ with factorisation-scale dependent parton distribution functions (PDFs)  $f_{i/h}(x_i,\mu_F^2)$ for a particle $i$ of momentum fraction $x_i$ in a hadron $h$,
\begin{equation}
	\label{eq:sigma}
	\begin{split}
	\sigma_{AB} = \int  M^2 \frac{\dd \sigma_{AB}}{\dd M^2}(\tau)  
	=& \sum_{a,b} \int_0^1 \dd{x_a} \dd{x_b} \dd{z} \Big[x_a f_{a/A}(x_a,\mu_F^2)\Big]\Big[x_b f_{b/B}(x_b,\mu_F^2)\Big]
	\\ \times &\Big[z \dd \sigma_{ab}(z,M^2,\mu_R^2,\mu_F^2)\Big]\ \delta(\tau - x_a x_b z) \,,
	\end{split}
\end{equation}
where $\tau = M^2/S$ is the ratio of the squared invariant mass $M^2$ over the hadronic centre-of-mass energy $S$ \cite{hep-ph/0409313}.

The partonic fraction $z=\tau/(x_ax_b) = M^2/s$ is defined by the ratio of the squared invariant mass to the partonic centre-of-mass energy $s=x_ax_bS$ and equals one at leading order. The partonic cross section
\begin{equation}
	\sigma_{ab}(s) = \int_2 \dd \sigma_{ab} = \int \frac{1}{2s}\ \overline{|\mathcal M|^2}\ \dd \text{PS}^{(2)}
\end{equation}
is related to the squared and averaged matrix element $\overline{|\mathcal M|^2}$ by the usual flux factor $1/(2s)$ and the integration over the two-particle phase space dPS$^{(2)}$.

\begin{figure}
	  \centering
	  \includegraphics[height=.25\linewidth]{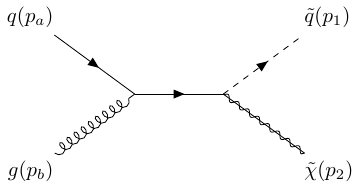}
	  \quad
	  \includegraphics[height=.25\linewidth]{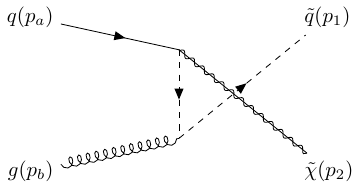}
	\caption{Tree-level $s$- (left) and $u$-channel (right) Feynman diagrams for the associated production of a squark and an electroweakino at hadron colliders.}
	\label{fig:lo_feyn}
\end{figure}

The associated production of a squark and an electroweakino with masses $m_\squark$ and $m_\gaugino$ occurs at a hadron collider at LO through the annihilation of a (massless) quark and a gluon. Charge conservation restricts the possible partonic processes to
\begin{align}\label{eq:processes}
	\quark_{u,d}(p_a)\ \gluon(p_b) \to \squark_{u,d} (p_1)\ \gaugino^0_{k} (p_2)\qquad\text{and}\qquad
	\quark_{u,d}(p_a)\ \gluon(p_b) \to \squark_{d,u}\ (p_1)\gaugino^\pm_{k} (p_2)
	\,,
\end{align}
where $k$ identifies the neutralino ($\gaugino^0_{k}$, $k=1,\dots,4$) or chargino ($\gaugino^\pm_{k}$, $k=1,2$) mass eigenstate and $p_{a,b}$ and $p_{1,2}$ refer to the four-momenta of the initial- and final-state particles, respectively. The corresponding Born diagrams are shown in \cref{fig:lo_feyn}. The squared matrix elements associated with the $s$-channel quark exchange diagram (left), the $u$-channel squark exchange diagram (right) and  their interference can be expressed as functions of the usual Mandelstam variables $s=(p_a+p_b)^2$, $t=(p_a-p_1)^2$ and $u=(p_a-p_2)^2$, {\it i.e.} as \cite{Dawson:1983fw}
\begin{align}
|\mathcal M_s |^2  &= \frac{g_s^2 C_A C_FB}{ s}\  2 (m_\gaugino^2-t) \,,\\
|\mathcal M_u |^2  &= - \frac{g_s^2 C_A C_FB}{ (u-m_{\tilde q}^2)^2}\ 2 (m_\chi^2-u)(m_{\tilde q}^2+u)\qquad {\rm and}\\
2 \Re[\mathcal M_s \mathcal M_u^\dagger] &=2\frac{g_s^2 C_AC_FB}{ s(u-m_\squark^2)}  
\Big(2(m_\gaugino^4-m_\squark^4) + m_\squark^2(2u-3s) -2m_\gaugino^2 ( 2m_\squark^2 + u)-su\Big)\,.
\end{align}
They are all proportional to the squared electroweakino-squark-quark coupling
\begin{equation}
	B\equiv R_{Ijk} L'_{Ijk}+L_{Ijk}R'_{Ijk} =R_{Ijk} R_{Ijk}^*+L_{Ijk}L_{Ijk}^* = |R_{Ijk}|^2 + |L_{Ijk}|^2
	\,,
\end{equation}
where the capitalised index $I$ labels the quark generation, the lower-case index $j$ refers to the squark eigenstate and the index $k$ is related as above to the electroweakino eigenstate. The definitions of the various left- and right-handed couplings $L^{(')}$ and $R^{(')}$ are provided in Refs.~\cite{Haber:1984rc,Gunion:1984yn,1604.01023}. Using arbitrary squark mixings and mass eigenstates opens the possibility to study SUSY flavour violation~\cite{Bozzi:2005sy,Bozzi:2007me, Fuks:2008ab,Fuks:2011dg,DeCausmaecker:2015yca,Chakraborty:2018rpn}.
The total spin- and colour-averaged squared amplitude then reads
\begin{equation}
	\overline{|\mathcal M|^2} = \frac{1}{96} \left(|\mathcal M_s |^2  +|\mathcal M_u |^2 + 2 \Re[\mathcal M_s \mathcal M_u^\dagger]\right)\,.
\end{equation}

The NLO corrections to this cross section are well-known \cite{1108.1250,1907.04898,2110.04211}.
They involve one-loop self-energy, vertex and box corrections interfering with tree-level diagrams, as well as squared real gluon and quark emission diagrams, from which intermediate on-shell squark and gluino resonant contributions have to be subtracted to avoid spoiling the predictivity of the NLO calculation and double-counting contributions to squark-pair production and gluino-electroweakino associated production with the corresponding subsequent decays~\cite{Gavin:2013kga,Hollik:2012rc}. We have calculated the full NLO cross section using dimensional regularisation of ultraviolet (UV) and infrared (IR) divergences as well as on-shell renormalisation for all squark and gluino masses and wave functions. The strong coupling constant is renormalised in the five-flavour $\overline{\text{MS}}$ scheme after explicitly decoupling the heavier coloured particles from its running~\cite{Collins:1978wz,Bardeen:1978yd,Marciano:1983pj}, which leaves the running determined only by the lightest coloured particles as it is usually done in global determinations of PDFs. In order to avoid the violation of SUSY invariance by the introduction of a mismatch between the strong coupling $g_s$ and the quark-squark-gluino Yukawa coupling $\hat g_s$ at one loop, we shift $\hat g_s$ by a finite contribution, allowing us to restore SUSY and to make use of the above-mentioned renormalisation scheme~\cite{hep-ph/9308222}.

Real and virtual contributions are combined with the help of the dipole subtraction method to cancel infrared and collinear divergences~\cite{hep-ph/9605323,hep-ph/0201036,hep-ph/0011222}. This method splits the pure NLO cross section into separately finite virtual and real contributions and a collinear counterterm consisting of two insertion operators $\mathbf P$ and $\mathbf K$,
\begin{equation}
	\label{eq:nlo}
\begin{split}
	\sigma^\text{NLO} &= \int_{3} \Big[ \dd\sigma^{\text{R}} - \dd\sigma^{\text{A}}\Big]_{\epsilon=0} 
	+ \int_{2} \bigg[\dd\sigma^{\text{V}} + \int_1\dd\sigma^{\text{A}}\bigg]_{\epsilon=0} \\
	&\qquad + \int_0^1\dd x \int_2 \Big[\dd\sigma^\text{B}(xp) \otimes (\mathbf P + \mathbf K)(x)\Big]_{\epsilon=0} 
	\,.
\end{split}
\end{equation}
In this expression the integration domain denotes the number of final-state particles. Moreover, the auxiliary cross section $\sigma^A$ shifts the infrared divergences such that the integrations over both the two- and three-particle phase spaces are numerically possible without changing the total result.

\subsection{Refactorisation}
\label{sec:ref}

After the cancellation of soft and collinear divergences among the real and virtual corrections, large logarithms remain near threshold~\cite{Kinoshita1962,Lee1964}. They arise between the constrained integration over the real emission phase space and the integration of virtual loops and take the form
\begin{equation}
	\left( \frac {\alpha_s} {2 \pi}\right)^n \left[ \frac{\log^m(1-z)}{1-z}\right]_+ \,,
\end{equation}
relative to the Born cross section with $m\leq 2n -1$. The variable $1-z = 1- M^2/s$ describes the energy fraction of an additional emitted gluon or massless quark and thus quantifies the distance to the partonic threshold. For soft emitted particles ($z \to 1$), truncating the perturbative calculation at a fixed order does not give a reliable prediction, so that the logarithms must be resummed to all orders in $\alpha_s$.

To calculate soft gluon emission up to all orders, kinematic and dynamical factorisation are necessary. Kinematic factorisation is possible by transforming the constituents of \cref{eq:sigma} into Mellin space
\begin{equation}
	\tilde{F}(N) = \int_0^1 \dd{y} y^{N-1} F(y)\,,
\end{equation}
with $F=\sigma_{AB},\, \sigma_{ab},\, f_{a/A},\, f_{b/B}$ and $y=\tau,\, z,\, x_a,\, x_b$, respectively. Denoting in the following all quantities in Mellin space and therefore dropping the tilde for simplicity, we obtain
\begin{equation}
	M^2 \frac{\dd \sigma_{AB}}{\dd M^2} (N-1) = \sum_{a,b}\ f_{a/A}(N,\mu_F^2)\ f_{b/B}(N,\mu_F^2)\  \sigma_{ab}(N,M^2,\mu_F^2,\mu_R^2)\, ,
\label{eq:HadFacN}
\end{equation}
such that the phase space factorises. In this expression, the large logarithms now depend on the Mellin variable $N$. Dynamical factorisation can then be achieved by relying on eikonal Feynman rules. The partonic cross section can be refactorised and resummed to 
\begin{equation}
	\begin{split}
	\sigma_{ab\to ij}(N,M^2,\mu_F^2,\mu_R^2) &= \sum_I \mathcal H_{ab \to ij,I}(M^2,\mu_F^2,\mu_R^2)\ \Delta_a(N,M^2,\mu_F^2,\mu_R^2)	\\
	&\qquad  \times \Delta_b(N,M^2,\mu_F^2,\mu_R^2)\ \Delta_{ab\to ij,I}(N,M^2,\mu_F^2,\mu_R^2)\,,
	\end{split}\label{eq:HtimesG}
\end{equation} 
in which the hard function is given by
\begin{equation}
	\mathcal H_{ab\to ij,I}(M^2,\mu_F^2,\mu_R^2) = \sum_{n=0}^\infty \left(\frac{\alpha_s}{2 \pi}\right)^n \mathcal H^{(n)}_{ab \to ij,I}(M^2,\mu_F^2,\mu_R^2)
	\,.
\end{equation}
This quantity is further discussed in \cref{sec:hard}~\cite{hep-ph/0409313,Kidonakis:1997gm,hep-ph/9801268}. 
The irreducible colour representation index $I$ is dropped from now on, since squark-electroweakino associated production involves only a single colour tensor. The soft wide-angle function $\Delta_{ab\to ij}$ and the soft collinear radiation functions $\Delta_{a,b}$ exponentiate~\cite{Sterman:1986aj,Catani1989,Catani1991,hep-ph/0010146},
\begin{equation}
	\label{eq:exponentiation}
	\Delta_a \Delta_b \Delta_{ab \to ij} = \exp\left[ LG^{(1)}_{ab}(\lambda) + G^{(2)}_{ab \to ij}(\lambda,M^2,\mu_F^2,\mu_R^2) + \dots \right]\,,
\end{equation}
with $\lambda = \alpha_s\beta_0L/(2\pi)$, $L = \log{} \bar N$ and $\bar N = Ne^{\gamma_E}$. The above expression contains the leading-logarithmic $G_{ab}^{(1)}$ and next-to-leading logarithmic $G_{ab \to ij}^{(2)}$ contributions. They are given by
\begin{align}
	G^{(1)}_{ab}(\lambda) &= g_a^{(1)}(\lambda) + g_b^{(1)}(\lambda) \,,
	\\
	G^{(2)}_{ab \to ij}(\lambda,M^2,\mu_F^2,\mu_R^2) &= g_a^{(2)}(\lambda,M^2,\mu_F^2,\mu_R^2) + g_b^{(2)}(\lambda,M^2,\mu_F^2,\mu_R^2)  + h^{(2)}_{ab\to ij}(\lambda)\,,
\label{eq:sudakov}\end{align}
with
\begin{align}
	g_a^{(1)} &= \frac{A_a^{(1)}}{2 \beta_0 \lambda} \left[ 2 \lambda + (1- 2 \lambda) \log (1- 2\lambda)\right]
	\,,
	\\
	g_a^{(2)} &= \frac{A_a^{(1)} \beta_1}{2 \beta_0^3 } \left[ 2 \lambda + \log(1- 2 \lambda) + \frac 1 2 \log^2 (1- 2\lambda)\right] - \frac{A_a^{(2)}}{2 \beta_0^2} \left[ 2 \lambda + \log (1- 2 \lambda)\right]
	\nonumber\\&+ \frac{A_a^{(1)}}{2 \beta_0} \left[ \log (1- 2 \lambda ) \log\left(\frac{M^2}{\mu_R^2}\right) + 2 \lambda \log\left( \frac{\mu_F^2}{\mu_R^2}\right)\right] \,.
\end{align}
The resummation coefficients entering those quantities are
\begin{align}
	A_a^{(1)} = 2 C_a \qquad\text{and}\qquad
	A_a^{(2)} = 2 C_a \left[ \left( \frac{67}{18} - \frac{\pi^2}{6} \right) C_A - \frac{5}{9} n_f\right]\,,
\end{align}
with $C_a=C_F$ for quarks and $C_a=C_A$ for gluons. The last term in \cref{eq:sudakov} consists of the process-dependent contributions related to large-angle soft-gluon emissions. It reads
\begin{align}
	h^{(2)}_{ab \to ij}(\lambda) &= \frac{\log{(1-2\lambda)}}{2 \beta_0} D^{(1)}_{ab \to ij}
      =  \frac{\log{(1-2\lambda)}}{2 \beta_0} \frac{2\pi}{\alpha_s} \Re(\bar\Gamma_{ab\to ij})\,.
\end{align}
The one-loop coefficient $D_{ab \to ij}^{(1)}$ does not vanish for squark-gaugino production, since soft gluons can be radiated off the final-state squark, and it depends on the modified soft anomalous dimension 
\begin{equation}
	\label{eq:soft:a:dim}
	\bar{\Gamma}_{\quark \gluon \rightarrow \squark \gaugino} 
	 ={} \frac{\alpha_s}{2\pi}\bigg\{ C_F \left[ 2\log\left(\frac{m_{\tilde{q}}^2 - t}{\sqrt{s} m_{\tilde{q}}}\right) - 1 + i\pi \right] + C_A  \log\left(\frac{m_{\tilde{q}}^2 - u}{m_{\tilde{q}}^2 - t}\right) \bigg\}\,,
\end{equation}
that has been derived in \cref{app:soft}.

\subsection{Hard matching coefficient}
\label{sec:hard}

The resummation of the logarithmic contributions as performed in \cref{eq:exponentiation} scales with the hard function $\mathcal H_{ab \to ij}(M^2,\mu_F^2,\mu_R^2)$, as shown in \cref{eq:HtimesG}. Including higher-order contribution in the hard function hence further improves the accuracy of the predictions. Therefore, in addition to the LO term
\begin{equation}
	\mathcal H^{(0)}_{ab \to ij}(M^2,\mu_F^2,\mu_R^2) = \sigma^{(0)}_{ab\to ij}(M^2)
	\,,
\end{equation}
we include the $N$-independent parts of the NLO cross section in the one-loop hard matching coefficient,
\begin{equation}
	\mathcal H^{(1)}_{ab \to ij}(M^2,\mu_F^2,\mu_R^2) = \sigma^{(0)}_{ab\to ij}(M^2)\ C^{(1)}_{ab\to ij}(M^2, \mu_F^2, \mu_R^2)
	\,.
\end{equation}

To compute this coefficient, we begin with the full NLO cross section of \cref{eq:nlo}. We first neglect the real emission contributions due to the three-particle phase space suppression close to threshold~\cite{Beenakker:2013mva,Beenakker:2011sf}. The virtual contributions $\dd\sigma^\text{V}$ and the integrated dipoles $\int_1 \dd \sigma^\text{A}$ in \cref{eq:nlo} correspond to a contribution proportional to $\delta(1-z)$, that is thus constant in $N$ after a Mellin transform. The collinear remainder is split into two pieces related to the insertion operators $\bf P$ and  $\bf K$, in which only the former depends on the factorisation scale $\mu_F$ \cite{hep-ph/9605323,hep-ph/0201036,hep-ph/0011222}. While logarithmic, but formally suppressed $\mathcal O(1/N)$ contributions have been shown to exponentiate and improve the numerical scale dependence in Drell-Yan like processes \cite{Kramer:1996iq,Harlander:2001is,Catani:2001ic}, we refrain from including them for our process. After discarding the $1/N$ terms that vanish in the large-$N$ limit, only the diagonal terms survive. We obtain for the initial quark
\begin{equation}\label{eq:k:quark}\begin{split}
 \Big\langle{\bf P}(N)\Big\rangle_q =&\ \frac{\alpha_s}{2\pi}\left(\log\bar{N} - \frac{3}{4}\right)\left(2 C_F \log\frac{\mu_F^2}{m_{\tilde{q}}^2 - t} - C_A \log\frac{s}{{m_{\tilde{q}}^2 - t}}\right)\,, \\
 \Big\langle{\bf K}(N)\Big\rangle_q =&\ \frac{\alpha_s}{2\pi}\Bigg\{ C_F \left(2\log^2\bar{N} + \frac{\pi^2}{2} - \frac{\gamma_q}{C_F} - \frac{K_q}{C_F}\right) \\ 
	  &+ \left(C_F - \frac{C_A}{2}\right)\left[2\log\bar{N}\left(1+\log\frac{m_\squark^2}{m^2_{\squark} - t}\right) + \mathcal{Q}\right]\Bigg\}\,,
\end{split}\end{equation}
with 
\begin{equation}
	\begin{aligned}
	 \mathcal{Q} ={}& \frac{m^2_{\squark} - t}{2m^2_{\squark} - t } + \frac{3 m_\squark}{m_\squark + \sqrt{2m^2_{\squark} - t }} + \log\frac{m_\squark^2}{2m^2_{\squark} - t } \left(1 + 2\log\frac{m_\squark^2}{m^2_{\squark} - t}\right) \\ 
	 &-\frac{3}{2}\log\frac{3m^2_{\squark} - t - 2m_\squark\sqrt{2m^2_{\squark} - t }}{m^2_{\squark} - t} + 2{\rm Li}_2 \left(\frac{2m^2_{\squark} - t}{m_\squark^2}\right) - \frac{\gamma_{\tilde{q}}}{C_F}
	\,.
	\end{aligned}
\end{equation}
For the initial gluon, we find
\begin{equation}\label{eq:k:gluon}\begin{split}
 \Big\langle{\bf P}(N)\Big\rangle_g =&\ \frac{\alpha_s}{2\pi}\left(C_A \log\bar{N} - \frac{\beta_0}{2}\right)\log\frac{\mu_F^4}{s(m_{\tilde{q}}^2 - u)} \,,\\
 \Big\langle{\bf K}(N)\Big\rangle_g =&\ \frac{\alpha_s}{2\pi}\frac{C_A}{2}\left[ 4\log^2\bar{N} \!+\! 2\log\bar{N}\left(1\!+\!\log\frac{m_\squark^2}{m^2_{\squark} \!-\! u}\right) \!+\! \pi^2 \!-\! \frac{2\gamma_g}{C_A} \!-\! \frac{2K_g}{C_A} \!+\! \mathcal{G}\right]\,,
\end{split}\end{equation}
with
\begin{equation}
      \begin{split}
	\mathcal{G} ={}& \frac{m^2_{\squark} - u}{2m^2_{\squark} - u } + \log\frac{m_\squark^2}{2m^2_{\squark} - u }\left(1 + 2 \log\frac{m_\squark^2}{m^2_{\squark} - u}\right) + 2{\rm Li}_2 \left(\frac{2m^2_{\squark} - u}{m_\squark^2}\right) - \frac{ \gamma_{\tilde q}}{C_F}\\
	&+
       \frac{\beta_0}{C_A} \left(\log\frac{3m^2_{\squark} - u-2m_\squark\sqrt{2m^2_{\squark} - u}}{m^2_{\squark} - u} + \frac{2m_\squark}{\sqrt{2m^2_{\squark} - u}+m_\squark}\right)
       \,.
       \end{split}
\end{equation}
In all these expressions, the two-body phase space Mandelstam variables are defined according to the particle ordering of \cref{eq:processes}, and the various constants are~\cite{hep-ph/0201036}
\begin{equation}
    \begin{aligned}
 &\gamma_q = \frac{3}{2}C_F\,, \qquad &&K_q = \left(\frac{7}{2}-\frac{\pi^2}{6}\right)C_F \,,\\ 
 &\gamma_g = \beta_0 = \frac{11}{6}C_A - \frac{2}{3}T_R N_f \,,\qquad &&K_g = \left(\frac{67}{18}-\frac{\pi^2}{6}\right)C_A - \frac{10}{9}T_R N_f \,,\\ 
 &\gamma_{\tilde{q}} = 2 C_F \,,\qquad &&K_{\tilde{q}} = \left(4-\frac{\pi^2}{6}\right)C_F 
    \,.
    \end{aligned}
\end{equation}
With the $N$-independent parts of the insertion operators, the hard functions read
\begin{eqnarray}
\mathcal H^{(0)}_{ab \to ij}(M^2,\mu^2) &=& \frac{\sigma^\text{B}(M^2)}{M^2}
\,,
\\
\mathcal H^{(1)}_{ab \to ij}(M^2,\mu^2) &=& \frac{2\pi}{\alpha_s} \frac{\sigma^\text{B}(M^2)}{M^2} \Big(\Big\langle{\bf P}+{\bf K}\Big\rangle_q+\Big\langle{\bf P}+{\bf K}\Big\rangle_g \Big)_{N\text{-ind.}}\nonumber
\\&+& \frac{2\pi}{\alpha_s}\Big(\frac{\sigma^\text{V}(M^2)+ \int_1 \dd\sigma^\text{A}(M^2)}{M^2} \Big) 
\,.
\end{eqnarray}
As we have ignored any $1/N$ terms in the above computation of the hard matching coefficient, we employ the standard collinear~unimproved resummation formalism as opposed to the collinear improved one of Refs.~\cite{Kramer:1996iq,Catani:2001ic,Kulesza:2002rh,Almeida:2009jt}. While only the $N$-independent terms are necessary in practice, we included the logarithmic terms in the above expressions to be able to validate analytically the re-expansion of the resummed cross section at ${\cal O}(\alpha_s^2)$ in \cref{sec:match}.

\subsection{Matching and expansion}
\label{sec:match}

So far we have computed a fixed order cross section $\sigma^\text{NLO}$ and a resummed cross section $\sigma^\text{Res.}$. As the latter is a good approximation near threshold and the former far from it, they should be consistently combined. Therefore, we sum up both contributions and remove the terms that are accounted for both in the resummed and the fixed-order predictions, thus avoiding any double counting. A consistent matching is achieved by re-expanding $\sigma^\text{Res.}$ at ${\cal O}(\alpha_s^2)$ and subtracting this quantity $\sigma^\text{Exp.}$ from the sum of the resummed and fixed-order results,
\begin{equation}
	\sigma_{ab} = \sigma_{ab}^\text{NLO} + \sigma_{ab}^{\text{Res.}}  - \sigma_{ab}^\text{Exp.}\,.
\end{equation}
The expansion is given in terms of the first- ($\mathcal H ^{(0)}$) and second-order ($\mathcal H^{(1)}$) hard function coefficients of \cref{sec:hard},
\begin{equation} 
	\label{eq:exp}
	\begin{split}
	\sigma^\text{Exp.}_{ab} &=   \mathcal H^{(0)}_{ab\to ij}(M^2,\mu^2) + \frac{\alpha_s}{2 \pi}\mathcal H^{(1)}_{ab\to ij}(M^2,\mu^2)+ \frac{\alpha_s}{2 \pi}\mathcal H^{(0)}_{ab\to ij}(M^2,\mu^2) \\
	&\times    \left( \Big(A_a^{(1)} + A_b^{(1)}\Big) \Big( \log \bar N + \log \frac {\mu_F^2}{M^2}  \Big)  - 2 D_{ab\to ij}^{(1)}\right) \log \bar N\,.
	\end{split}
\end{equation}
We can verify that the leading logarithmic terms in $\log^2 \bar N $ agree with those of the $\mathbf K$-operators for quarks ($A_q^{(1)}=2C_F$) and gluons ($A_g^{(1)}=2C_A$) in \cref{eq:k:quark} and \cref{eq:k:gluon}.
Moreover, combining the next-to-leading logarithmic terms in $\log \bar N$ originating from the soft anomalous dimension in \cref{eq:soft:a:dim} and the terms in $\log (\mu_F^2/s)$ from \cref{eq:exp}, we recover the same terms as in the sum of the contributions of the  $\mathbf P$ and $\mathbf K$ operators for quarks and gluons,
\begin{eqnarray}
 \left[\frac{\sigma^\text{Exp.}_{ab}M^2}{\sigma^\text{B}}\right]_{\log \bar N} &=&\left[\Big\langle{\bf P}(N)\Big\rangle_q+\Big\langle{\bf K}(N)\Big\rangle_q +\Big\langle{\bf P}(N)\Big\rangle_g+\Big\langle{\bf K}(N)\Big\rangle_g  \right]_{\log \bar N}  
\nonumber\\&=&
	2C_F  \left[ \log \frac{\mu_F^2 }{ m_\squark^2 -t} +\log \frac{ m_\squark^2}{ m_\squark^2 -t}  +1 \right]\log \bar N \\&+&
	2C_A \left[ \log \frac{\mu_F^2}{s}+ \log \frac{m_\squark^2-t}{m_\squark^2-u}\right]\log \bar N 
	\,.\nonumber
\end{eqnarray}

Having computed the resummed and the perturbatively expanded results
in Mellin space, we must multiply them with the $N$-moments of the PDFs according
to \cref{eq:HadFacN} and perform an inverse Mellin transform,
\begin{equation}
 M^2{\dd\sigma_{AB}\over\dd M^2}(\tau)={1\over2\pi i}\int_{{\cal C}_N}\dd N 
 \tau^{-N} M^2{\dd\sigma_{AB}(N)\over \dd M^2},
	\label{eq:inv:mass}
\end{equation}
in order to obtain the hadronic cross section as a
function of $\tau=M^2/S$. Special attention must be paid to the
singularities in the resummed exponents $G_{ab}^{(1,2)}$, which are situated at
$\lambda=1/2$ and are related to the Landau pole of the perturbative coupling
$\alpha_s$. To avoid this pole as well as those in the Mellin moments of the PDFs
related to the small-$x$ (Regge) singularity $f_{a/A}(x,\mu_0^2)\propto x^\alpha
(1-x)^\beta$ with $\alpha<0$, we choose an integration contour ${\cal C}_N$
according to the {\em principal value} procedure proposed in
Ref.~\cite{Contopanagos:1993yq} and the {\em minimal prescription} proposed in
Ref.~\cite{Catani:1996yz}. We define two branches
\begin{eqnarray}
  {\cal C}_N:~~ N=C+ze^{\pm i\phi}~~{\rm with}~~ z\in[0,\infty[,
  \label{eq:IT:Nbra}
\end{eqnarray}
where the constant $C$ is chosen such that the singularities of the $N$-moments
of the PDFs lie to the left and the Landau pole to the right of the
integration contour. Formally the angle $\phi$ can be chosen in the range
$[\pi/2,\pi[$, but the integral converges faster if $\phi>\pi/2$.

\section{Numerical Results}
\label{sec:num}

For our numerical predictions, we identify the Standard Model parameters with those determined by the Particle Data Group \cite{ParticleDataGroup:2020ssz}. The running of the strong coupling with five active quark flavours is chosen in agreement with the selected PDF set as provided by the \textsc{LHAPDF6} library \cite{lhapdf}. As our default choice of PDFs, we employ the sets of MSHT20~\cite{Bailey:2020ooq} unless stated otherwise. To be specific, we use at LO the set MSHT20LO130 with $\alpha_s(M_Z)=0.130$, and at NLO and NLO+NLL the set MSHT20NLO118 with $\alpha_s(M_Z)=0.118$. Again unless stated otherwise, we consider the dominant squark-electroweakino production channel at the LHC, {\it i.e.} the production of a left-handed or a right-handed up-type squark in association with the lightest neutralino.

\subsection{Validation of our NLO implementation}

In order to validate the numerical implementation of our NLO calculation, we compare our NLO predictions with those obtained with \textsc{MadGraph5$\_$aMC@NLO} \cite{1405.0301}, that rely on the MSSM model implementation described in Ref.~\cite{1907.04898}. In addition, we subtract on-shell squark and gluino contributions locally from the real emission pieces by reshuffling the momenta of the final-state particles and using a standard Breit-Wigner function for the decay \cite{1108.1250}. Technically, we employ the plugin \textsc{MadSTR} with the option \textsc{istr}$=5$ \cite{1907.04898}. We find excellent numerical agreement between the two approaches.

\begin{figure}
     \centering
     \begin{subfigure}[b]{0.49\textwidth}
         \centering
	  \includegraphics[width=\textwidth]{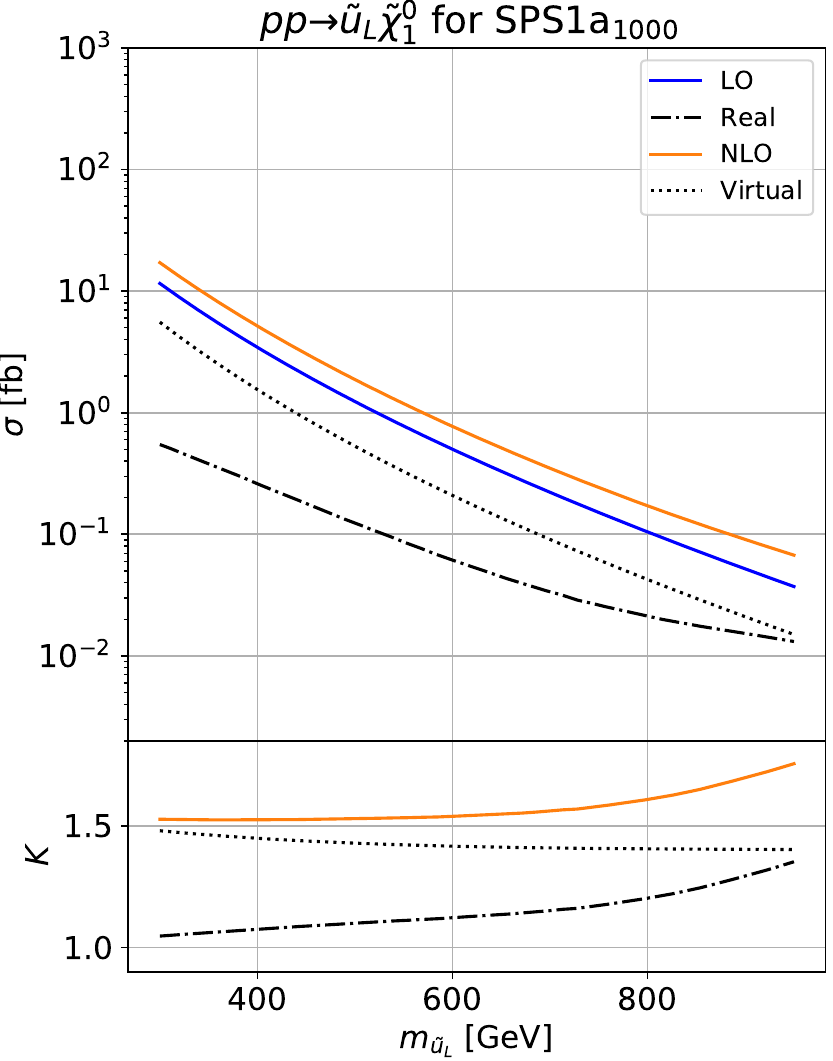}
     \end{subfigure}
     \hfill
     \begin{subfigure}[b]{0.49\textwidth}
         \centering
	  \includegraphics[width=\textwidth]{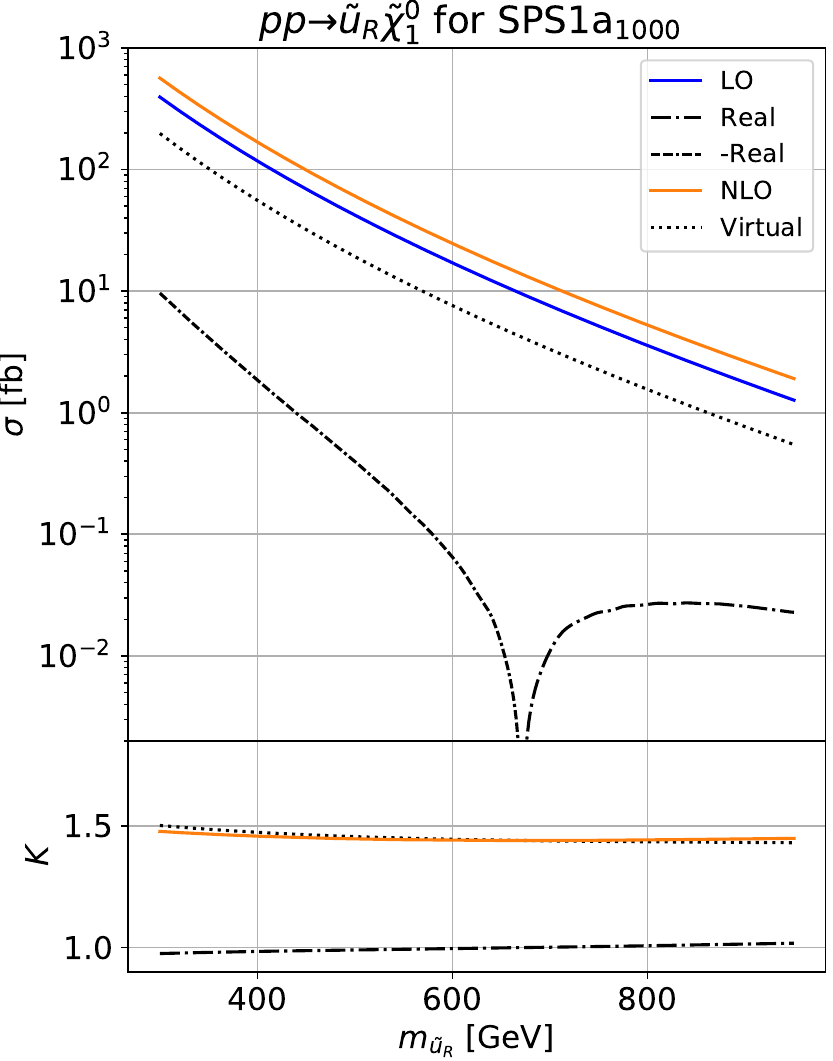}
     \end{subfigure}
     \caption{
    	NLO cross sections $\sigma(pp \to\tilde u_{L,R} \gaugino_1^0)$ and their virtual and real components (top panels) as well as the associated $K$ factors (bottom panels) as a function of the physical squark mass $m_{\upsquark_{L,R}}$ for a fixed mass difference $m_{\upsquark_L}-m_{\upsquark_R} = \SI{20}{GeV}$. This fixed mass difference is only adopted in order to reproduce Ref.\ \cite{1108.1250}, even though it breaks in general SU(2) invariance. The remaining MSSM parameters are fixed to the benchmark point SPS1a$_{1000}$, the LHC energy is $\sqrt{S}=\SI{7}{TeV}$, and CTEQ6.6M PDFs are employed.
     }
     \label{fig:mass_plehn}
\end{figure}

In addition, we validate our NLO implementation by studying the squark mass dependence of the total cross section and comparing our results to those of an independent, previously published automated calculation \cite{1108.1250} at the MSSM benchmark point SPS1a$_{1000}$. This benchmark is based on the point SPS1a \cite{hep-ph/0202233}, for which the physical particle spectrum is calculated with \textsc{SPheno} 3 \cite{Porod:2011nf}, and the gluino mass is then shifted to \SI{1}{TeV}. Our results are presented in \cref{fig:mass_plehn}.
The contributions from real and virtual corrections, made individually finite with the Catani-Seymour dipole formalism \cite{hep-ph/0201036}, are also shown separately. The contributions from the integrated dipoles and the collinear counterterms $\mathbf{P} + \mathbf{K}$ have been combined with the virtual corrections. For the associated production of a right-handed up-type squark with the lightest neutralino (right), we observe the same sign flip in the real corrections as the one that had been found in Ref.~\cite{1108.1250}. Moreover, our $K$-factors for the real and virtual corrections, defined by their ratios with respect to the Born cross section, exhibit the same behaviour as those in the reference computation, both for $\tilde u_L\tilde\chi_1^0$ and $\tilde u_R\tilde\chi_1^0$ production. The remaining minor numerical differences between the two calculations can be traced back to the use of a slightly different PDF set, as \textsc{LHAPDF6} \cite{lhapdf} no longer supports CTEQ6M~\cite{Pumplin:2002vw}, and benchmark scenario with small differences in the superpartner masses (below \SI{1}{\percent}).

\begin{figure}
\includegraphics[width=\linewidth]{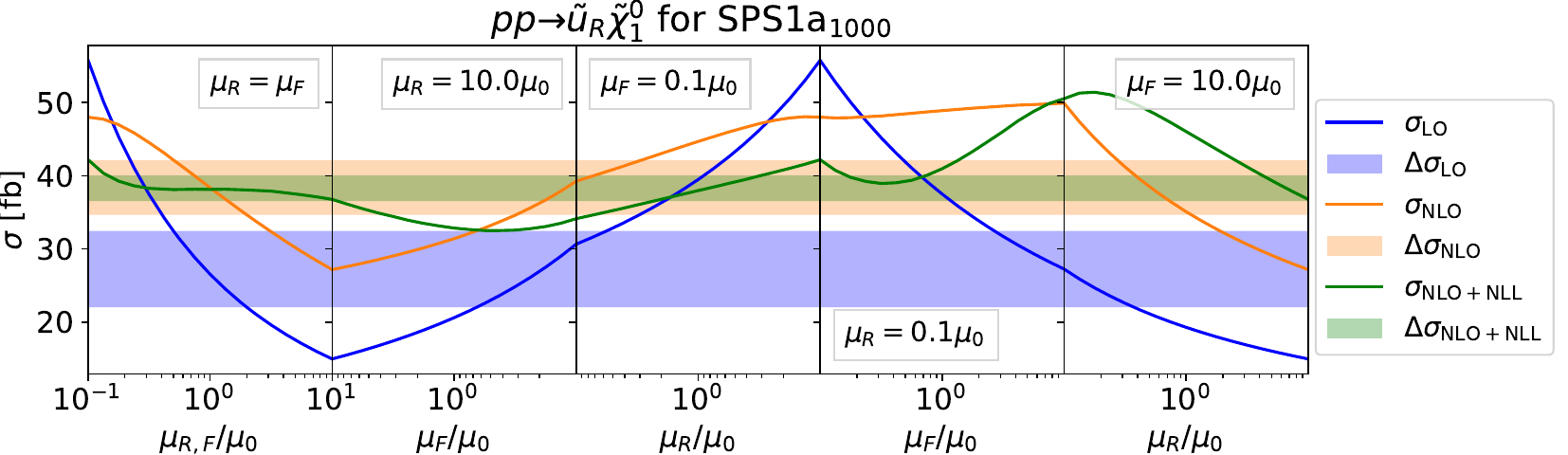}
\caption{
	Profile of the renormalisation and factorisation scale dependence of the total cross section for the process $pp \to\tilde u_{R} \gaugino_1^0$. The plot covers $\mu_{F,R}\in (0.1-10)\mu_0$ (reversely in panels 2 and 3), where the central scale is $\mu_0= (m_\squark + m_\gaugino)/2$. The bands show the scale uncertainties as obtained from the seven-point method. The predictions in this plot are for a centre-of-mass energy of $\sqrt{S}=$ 7 TeV and CTEQ6.6M PDFs.
}
\label{fig:scale_plehn}
\end{figure}

Next, we demonstrate in \cref{fig:scale_plehn}
that the factorisation and renormalisation scale dependence of our result agrees with the findings of Ref.~\cite{1108.1250}. In addition, we show as shaded bands the uncertainties obtained by varying the central scale $\mu_0\equiv\mu_R=\mu_F= (m_\squark + m_\gaugino)/2$ with the seven-point method, {\it i.e.} by varying both scales independently by a factor of two up and down, but excluding relative factors of four between the two scales. As expected, the NLO corrections increase the total cross section and reduce the scale uncertainties. For illustrative purposes, we include also our new NLL+NLO predictions (green). While they do not systematically increase the NLO cross section for the masses chosen here, the scale uncertainty is reduced by about a factor of two relative to the NLO result.

\subsection{Invariant mass distributions}

In the following, we present results for two specific phenomenological MSSM scenarios with eleven parameters (pMSSM-11). The input parameters and the relevant resulting physical masses obtained with {\sc SPheno} 3 are listed in \cref{tab:scenarios}.
\begin{table}
\begin{center}
	\begin{tabular}{ |c|c|c| } 
		\hline
		\multicolumn{3}{|c|}{pMSSM-11 scenario A }  \\
		\hline
		$M_1$ & $M_2$ & $M_3$     \\
		$0.25$ & $0.25$ & $-3.86$ \\
		\hline
 		$M_{(U,D,Q)_{1,2}}$ & $M_{(U,D,Q)_{3}}$& $\mu$\\
 		$4.0$ & $1.7$ & $1.33$\\		
		\hline
		 $M_{(L,E)_{1,2}}$ & $M_{(L,E)_{3}}$&$\tan \beta$ \\
		 $0.35$ & $0.47$ &$36$ \\
		\hline
		$M_A$ & $A_0$    &\\
		$4.0$ & $2.8$  & \\
		\hline
 		$m_{\gaugino_1^0}$  & $m_{\upsquark}$& $m_\gluino$\\
		 $0.249$  & $ 4.07 $ & $ 3.90 $\\
		\hline
	\end{tabular}\hspace{1cm}
	\begin{tabular}{ |c|c|c| } 
		\hline
		\multicolumn{3}{|c|}{pMSSM-11 scenario B }  \\
		\hline
		$M_1$ & $M_2$ & $M_3$    \\
		$0.51$ & $0.48$ & $3.00$ \\
		\hline
 		$M_{(U,D,Q)_{1,2}}$ & $M_{(U,D,Q)_{3}}$& $\mu$ \\
 		$0.9$ & $2.0$& $-9.4$ \\		
 		\hline
		 $M_{(L,E)_{1,2}}$ & $M_{(L,E)_{3}}$ &$\tan \beta$  \\
		 $1.85$ & $1.33$ &$33$ \\
		\hline
$M_A$ & $A_0$     &\\
$3.0$ & $-3.4$  &\\		
		\hline
		$m_{\gaugino_1^0}$& $m_{\upsquark}$& $m_\gluino$ \\
		 $0.505$ & $ 0.96 $& $ 2.94 $ \\
		\hline
	\end{tabular}
\end{center}
\caption{Higgs and soft SUSY breaking parameters in our pMSSM-11 benchmark models, together with the relevant resulting physical particle masses. All values, except for $\tan \beta$, are given in \si{TeV}.
        }
\label{tab:scenarios}
\end{table}
First, we focus on a scenario featuring large squark masses of \SI{4}{TeV}, referred to as scenario~A. Second, scenario~B explores squark and gaugino masses expected to be within the reach of Run 3 of the LHC. Both scenarios are based on the global fits of Ref.~\cite{1710.11091}. Scenario A is derived from a fit that includes data from the anomalous magnetic moment of the muon~\cite{Muong-2:2021ojo}, while scenario B does not include it. In addition, we have lowered the parameters $M_1$ and $M_2$ in scenario B and have increased the parameter $M_3$ to bring the squark and gluino masses in agreement with the current SUSY limits from the LHC \cite{2101.01629,2010.14293,1908.04722}.

\begin{figure}
     \centering
 \begin{subfigure}[b]{0.49\textwidth}
         \centering
	  \includegraphics[width=\textwidth]{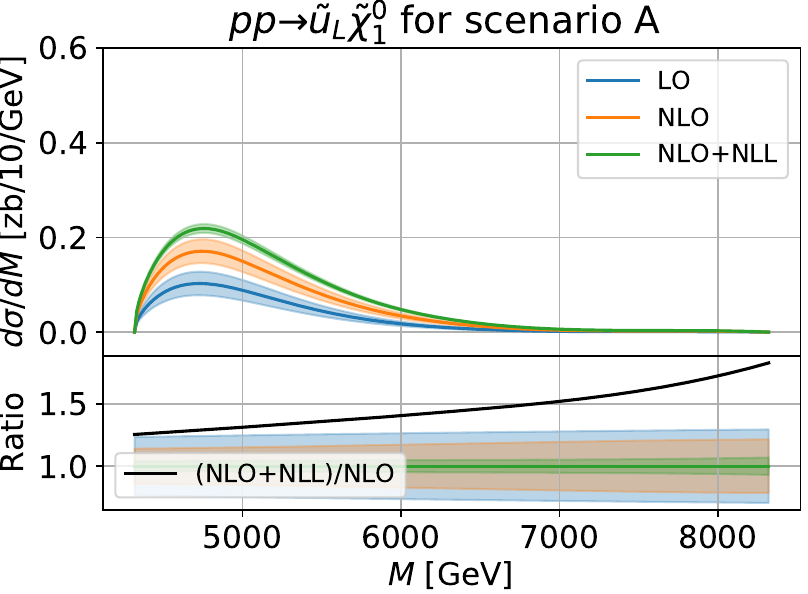}
     \end{subfigure}
     \hfill
     \begin{subfigure}[b]{0.49\textwidth}
         \centering
	  \includegraphics[width=\textwidth]{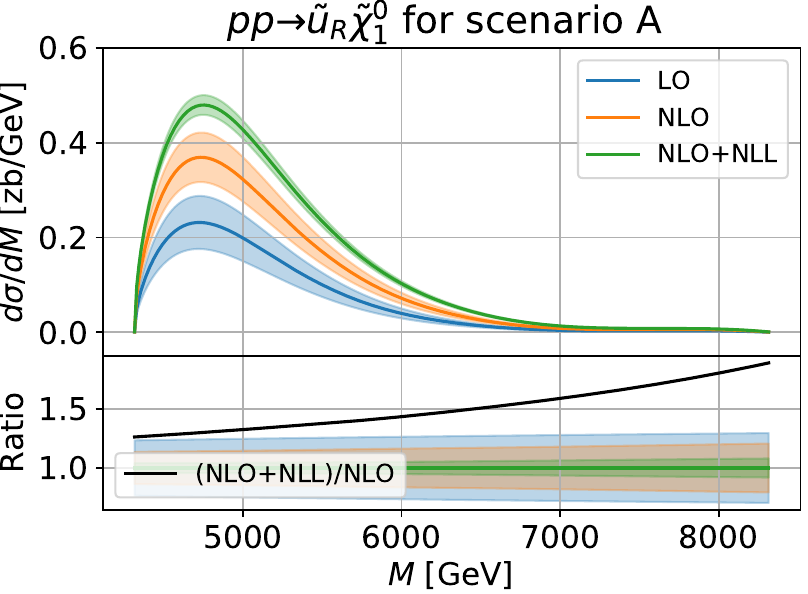}
     \end{subfigure}
     \newline
      \begin{subfigure}[b]{0.49\textwidth}
         \centering
	  \includegraphics[width=\textwidth]{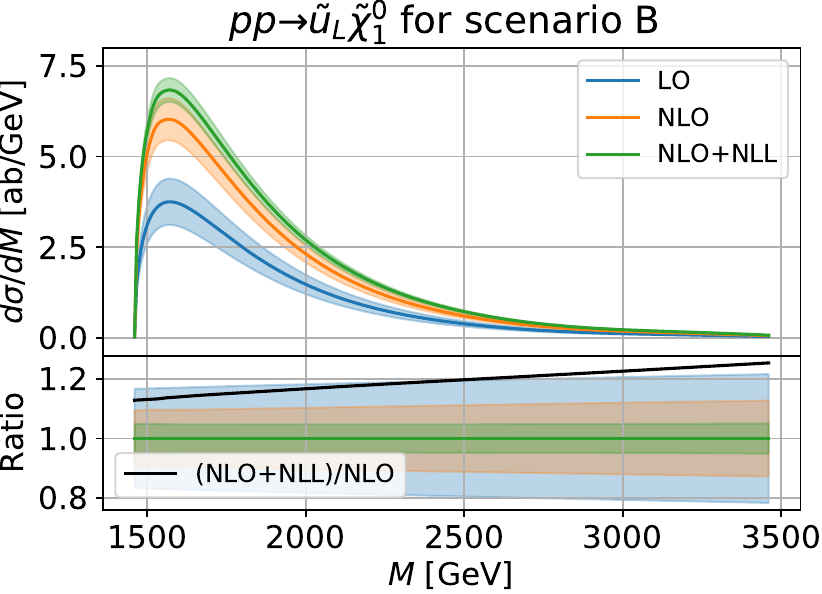}
     \end{subfigure}
     \hfill
     \begin{subfigure}[b]{0.49\textwidth}
         \centering
	  \includegraphics[width=\textwidth]{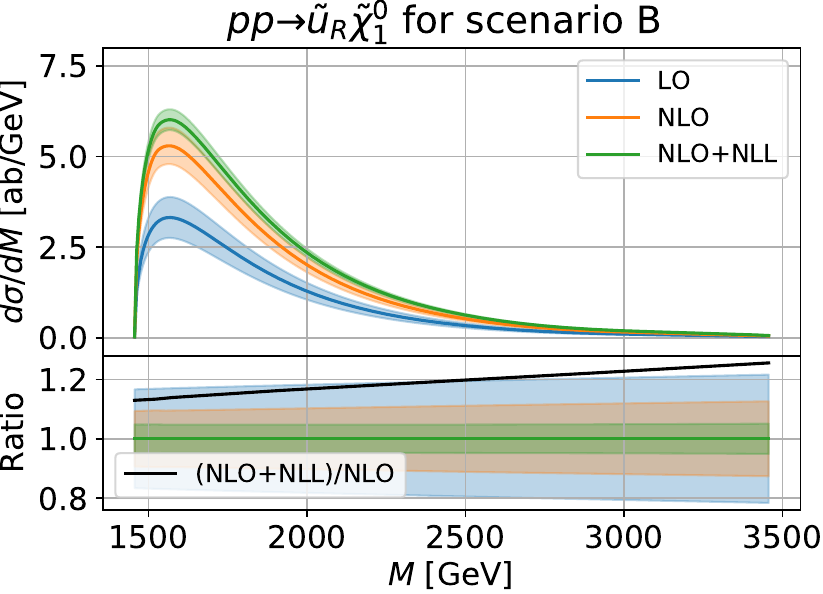}
     \end{subfigure}   
     \caption{Invariant-mass distributions for the processes $pp \to \upsquark_{L,R} \gaugino_1^0$ (top panels). The uncertainties correspond to variations around the central scale $\mu_0=M$, and we additionally show NLL+NLO/NLO $K$-factors (bottom panels). The results are shown for both scenarios A and B with MSHT20 PDFs and a centre-of-mass energy of $\sqrt{S}= \SI{13}{TeV}$.
     }
     \label{fig:invariant_mass}
\end{figure}

In \cref{fig:invariant_mass},
we show invariant mass distributions as derived from \cref{eq:inv:mass} for the associated production of left- and right-handed squarks with the lightest neutralino in both scenarios A and B. An invariant mass below the combined final state particle masses $M_0=m_\gaugino + m_\squark$ is kinematically forbidden. Very close to this lower limit there is a rapid increase of the cross section, until it peaks at about $M \approx 1.1 M_0$ and then falls off towards higher values of $M$. As the invariant mass $M$ increases, we get closer to the threshold region $z = M^2/s \to 1$, and NLL corrections contribute significantly more to the differential cross section.  This behaviour is captured by the NLL+NLO/NLO $K$-factors shown in the lower panels of the figure. For scenario A, the increase of the NLO cross section goes from \SI{25}{\percent} in the region of the peak to more than \SI{50}{\percent} at large invariant masses. In contrast, scenario B receives smaller corrections of \SI{10}{\percent} to \SI{20}{\percent} due to the smaller invariant mass $M$ relevant for the bulk of the cross section.

The lower panels display the relative seven-point scale uncertainty, when the scale is varied around a central scale choice of $\mu_0 = M$. Across the whole mass range, the scale uncertainty is first reduced when comparing LO rates to NLO ones, and next further reduced when considering NLO+NLL predictions. By performing a similar calculation with $\mu_0 = (m_\squark + m_\gaugino)/2$ we recover the total cross section (see below) as the area under the invariant mass distribution.

In our figures, results for the production of left-handed up-squarks in scenario A are scaled by a factor of 10. Due to the bino-like nature of the neutralino $\gaugino^0_1$ in this scenario, the coupling to the right-handed up-squark is dominant, and so is the cross section related to the production of a $\tilde u_R\tilde\chi_1^0$ pair. In scenario B however, the composition of $\gaugino^0_1$ is roughly \SI{50}{\percent} wino and \SI{50}{\percent} bino, yielding cross sections of the same order for the two processes $pp\to \tilde u_R\tilde\chi_1^0$ and $pp\to \tilde u_L\tilde\chi_1^0$.

\subsection{Total cross sections and their scale uncertainty}

\begin{figure}
	\centering
	\begin{subfigure}[b]{\textwidth}
	    \centering
\includegraphics[width=\linewidth]{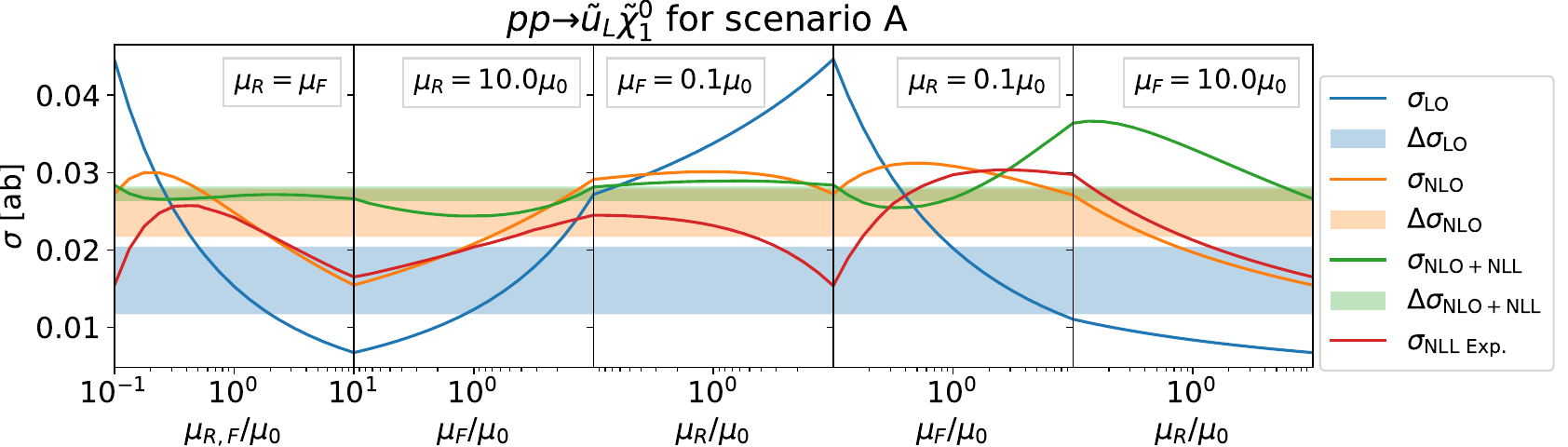}
	\end{subfigure}
	\newline
	\begin{subfigure}[b]{\textwidth}
	    \centering
\includegraphics[width=\linewidth]{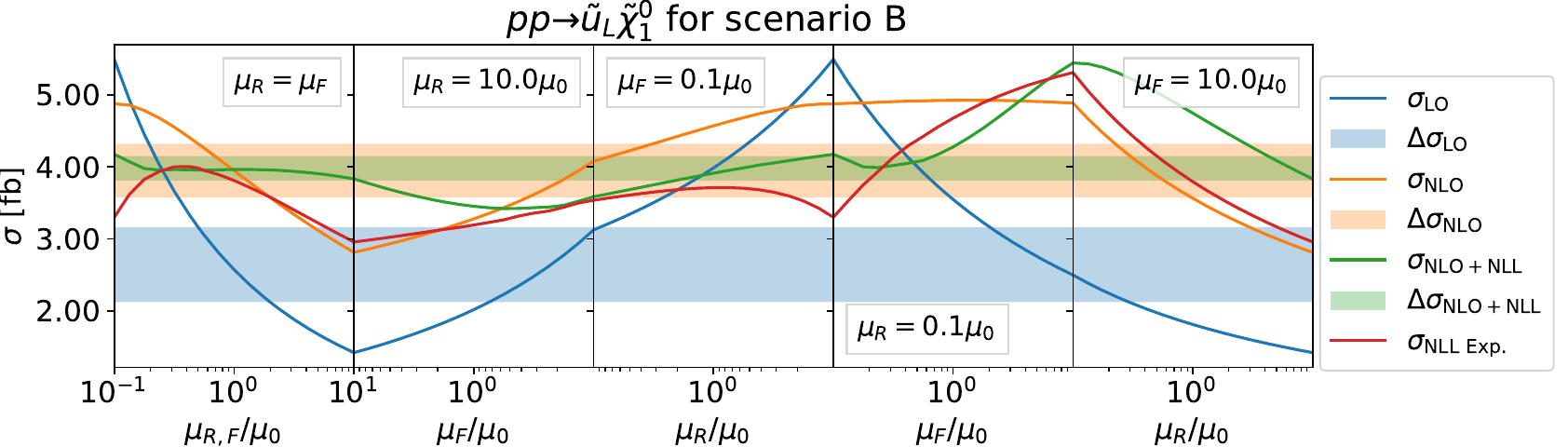}
	\end{subfigure}
	\caption{
	Profiles of the renormalisation and factorisation scale dependence  of the total cross section corresponding to the process $pp \to\tilde u_{L} \gaugino_1^0$ in scenarios A and B. The plots cover $\mu_{F,R} \in (0.1-10)\mu_0$ (reversely in panels 2 and 3) with a central scale $\mu_0= (m_\squark + m_\gaugino)/2$. The bands correspond to scale uncertainties evaluated with the seven-point method, and we use $\sqrt{S}=\SI{13}{TeV}$ and MSHT20 PDFs. We present predictions at LO, NLO and NLO+NLL, as well as for the ${\cal O}(\alpha_s^2)$ expansion of the NLL result.
	}
	\label{fig:scale}
 \end{figure}

In \cref{fig:scale},
we present predictions for the total cross section related to the process $pp\to\tilde u_L\tilde\chi_1^0$ in scenarios A and B with squark masses of \SI{4}{TeV} and 1 TeV, respectively, together with the associated scale uncertainties. The results are shown at LO, NLO and NLO+NLL for a centre-of-mass energy of $\sqrt{S}=13$~TeV. Our predictions show a significant increase of the total cross section in scenario A when including NLO+NLL corrections as well as a reduction of the scale uncertainties in both scenarios. The uncertainty bands are again determined by the seven-point method, where the factorisation and renormalisation scales are both varied independently by factors of two up and down around the central scale $\mu_0=(m_\squark+ m_\gaugino)/2$, excluding the cases where $\mu_{F}/\mu_{R} = 4$ or $1/4$.  For both examined scenarios, we observe that the relative scale uncertainties are reduced from about \SI{\pm 20}{\percent} at LO to \SI{\pm 10}{\percent} at NLO, and finally fall below \SI{\pm 5}{\percent} at NLO+NLL. The kink in the NLO cross section at $\mu_F=\mu_R=0.1\mu_0$ between panels three and four is more prominent in scenario A than in scenario B. It originates from the subtraction of on-shell squark and gluino resonant contributions from the real emission component of the cross section. We also include predictions for the expansion of the NLL predictions at ${\cal O}(\alpha_s^2)$, following \cref{eq:exp} (solid red curve). As expected for large scales the logarithmic terms become dominant, and the expansion consequently approximates well the NLO result. Full control over the scale dependence at next-to-next-to-leading order (NNLO) and beyond can of course only be obtained with an explicit calculation.

\subsection{Parton density uncertainties of the total cross section}

So far we have only studied scale uncertainties associated with the total rates for squark-electroweakino production at the LHC. There is, however, a second important source of theoretical uncertainties, {\it i.e.}~those coming from the parton density fits. To compute them, we use the methods available in \textsc{LHAPDF}, which allow to calculate  the PDF uncertainties in two different ways \cite{pdfmonte}:
\begin{description}
	\item[ - Eigenvectors:] Experimental uncertainties are parametrised by making computations for a set of orthogonal Hessian PDF eigenvectors.
	The uncertainty is calculated from  
	\begin{equation}
		\Delta\sigma_{\text{PDF}\pm} = \sqrt{\sum_{i=1}^n \Big[\max(\pm \sigma_{+i} \mp \sigma_0,\ \pm \sigma_{-i}\mp \sigma_0,0 )\Big]^2}\,,
	\end{equation}
	where the index $i$ runs over all PDF eigenvectors, with $i$ = 0 representing the central, best fit, set. This method is the one to be used with CT18~\cite{Hou:2019efy} and MSHT20~\cite{Bailey:2020ooq} densities.
	\item[ - Replicas:] Monte Carlo PDF sets are provided with multiple replicas that need to be combined to get a symmetric uncertainty on the predictions. This is achieved through the formula
	\begin{equation}
		\Delta\sigma_{\text{PDF}\pm} = \sqrt{\frac{1}{n-1}\sum_{i=1}^n \Big[\sigma_{i}-\langle\sigma\rangle\Big]^2}\,,
	\end{equation}
	where the index $i$ runs over the entire set of PDF replicas, with the central value being given by the mean cross section value $\langle \sigma \rangle = \frac{1}{n} \sum^n_{i=1} \sigma_i\simeq \sigma_0$.
	This method is used with NNPDF40 densities~\cite{Ball:2021leu}. 
\end{description}

\begin{figure}
     \centering
 \begin{subfigure}[b]{0.49\textwidth}
         \centering
	  \includegraphics[width=\textwidth]{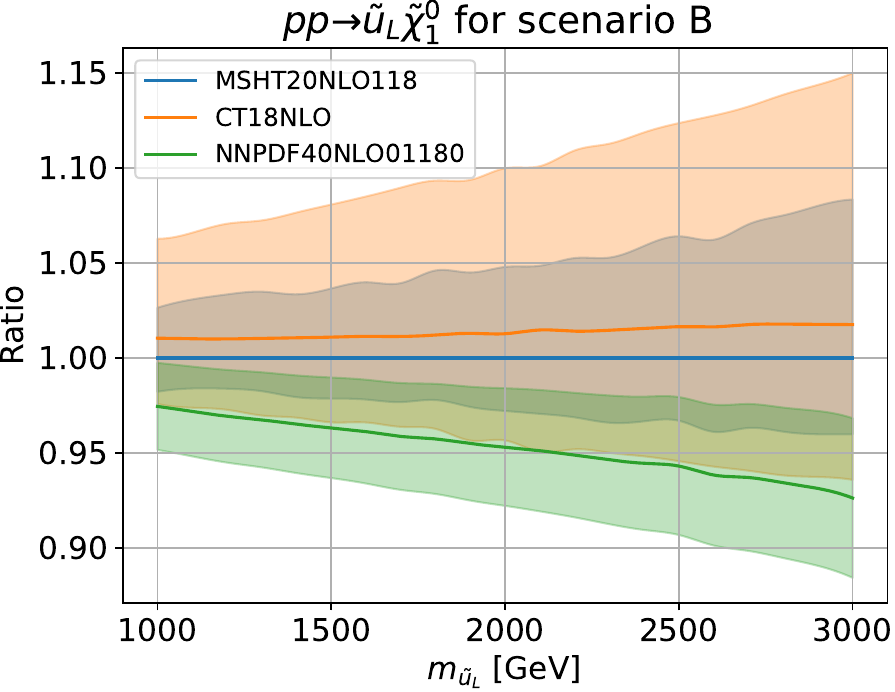}
     \end{subfigure}
     \hfill
     \begin{subfigure}[b]{0.49\textwidth}
         \centering
	  \includegraphics[width=\textwidth]{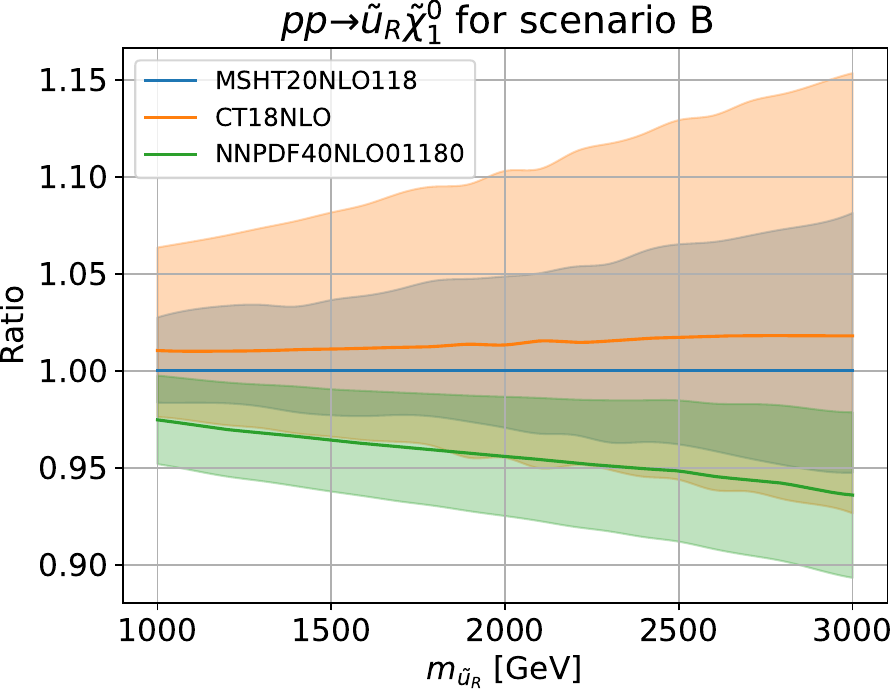}
     \end{subfigure}
     \newline
      \begin{subfigure}[b]{0.49\textwidth}
         \centering
	  \includegraphics[width=\textwidth]{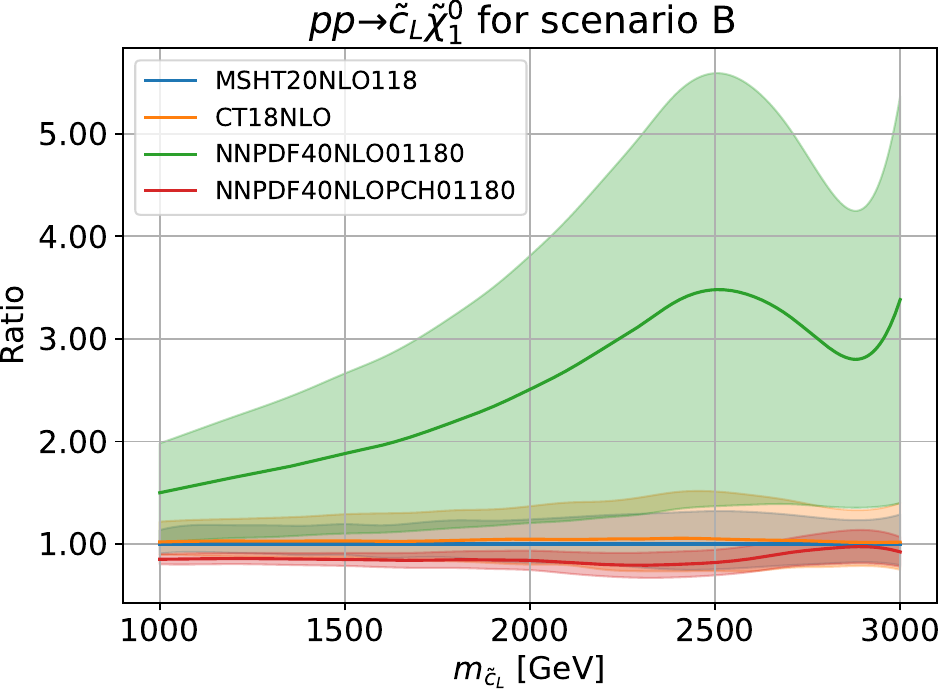}
     \end{subfigure}
     \hfill
     \begin{subfigure}[b]{0.49\textwidth}
         \centering
	  \includegraphics[width=\textwidth]{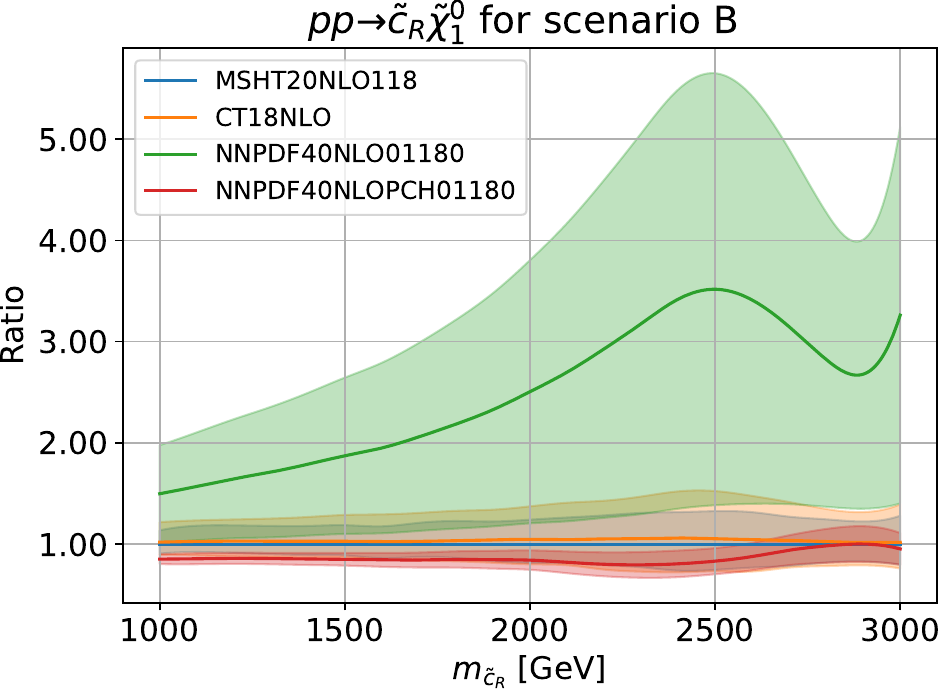}
     \end{subfigure}   
     \caption{
	    Relative PDF uncertainties of the total cross sections for the processes  $pp \to\tilde q_{L,R} \gaugino_1^0$ at the LHC with a centre-of-mass energy of $\sqrt{S}=\SI{13}{TeV}$ and at NLO+NLL. 
        The uncertainties are shown as a function of the squark mass $m_{\squark_{L,R}}$ and for four different choices of PDFs, namely MSHT20NLO118 (blue), CT18NLO (orange), NNPDF40NLO01180 (green) and NNPDF40NLOPCH01180 (red).
     }
     \label{fig:pdfs}
\end{figure}

For our predictions at NLO and NLO+NLL, we calculated the PDF uncertainties at \SI{90}{\percent} confidence level following the convention of Ref.\ \cite{Hou:2019efy}. The resummation of large logarithms does not significantly alter the size of the relative PDF uncertainties, as the same set of PDFs is used in both calculations. For scenario B we show in \cref{fig:pdfs} the PDF uncertainties associated with NLO+NLL total cross sections for the different choices of parton densities mentioned above, {\it i.e.}~for MSHT20, CT18 and NNPDF40. We consider the process $pp\to \tilde u_{L,R}\tilde\chi_1^0$ in the top row of the figure and present predictions as a function of the squark mass. While the uncertainty is of about \SI{5}{\percent} for \SI{1}{TeV} up-squarks, it increases up to $10-15$ \si{\percent} for squark masses of \SI{3}{TeV}. This increase is related to the large partonic momentum fractions $x$ relevant for such a large mass, where the PDFs are less constrained in their fitting procedure. The central cross section values obtained with the MSHT20 and CT18 sets agree consistently at the percent level in the explored mass range, the MSTH20 errors being slightly smaller as a consequence of this set being more recent than the CT18 one. On the other hand, the NNPDF40 predictions are a few percent lower, although they are still in reasonable agreement within their uncertainty intervals with the predictions achieved with other PDFs.

In the lower two plots of \cref{fig:pdfs} we show results for the production of a charm squark in association with the lightest neutralino. We observe that the results obtained with the central NNPDF40NLO01180 set with $\alpha_s(M_Z)=0.118$ give cross sections that are larger by a factor of three to four with respect to those obtained with the CT18NLO set and with the MSHT20NLO118 set, both also with $\alpha_s(M_Z)=0.118$. In addition, the uncertainties associated with the NNPF40 predictions are of about 30--50\si{\percent}, in contrast with predictions obtained with CT18 and MSHT20, that have much smaller uncertainties. This discrepancy can be traced back to the treatment of the charm quark in the NNPDF40 fit~\cite{2104.09174} and is expected to be even more significant in processes with two charm quarks or antiquarks in the initial state. The cross sections estimated with the alternative NNPDF40NLOPCH01180 PDF fit (red) with $\alpha_s(M_Z)=0.118$, in which the treatment of the charm quark is kept purely perturbative, are, in contrast, in good agreement with CT18 and MSHT20 both for the central values and the uncertainties.

\subsection{Squark and gaugino mass dependence of the total cross section}

\begin{figure}
	\centering
	\begin{subfigure}[b]{0.49\textwidth}
	    \centering
	     \includegraphics[width=\textwidth]{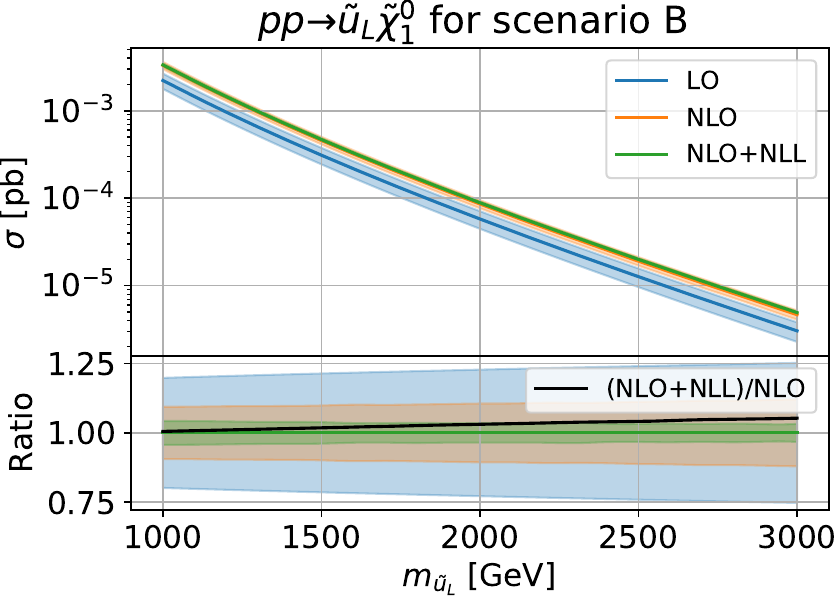}
	\end{subfigure}
	\hfill
	\begin{subfigure}[b]{0.49\textwidth}
	    \centering
	     \includegraphics[width=\textwidth]{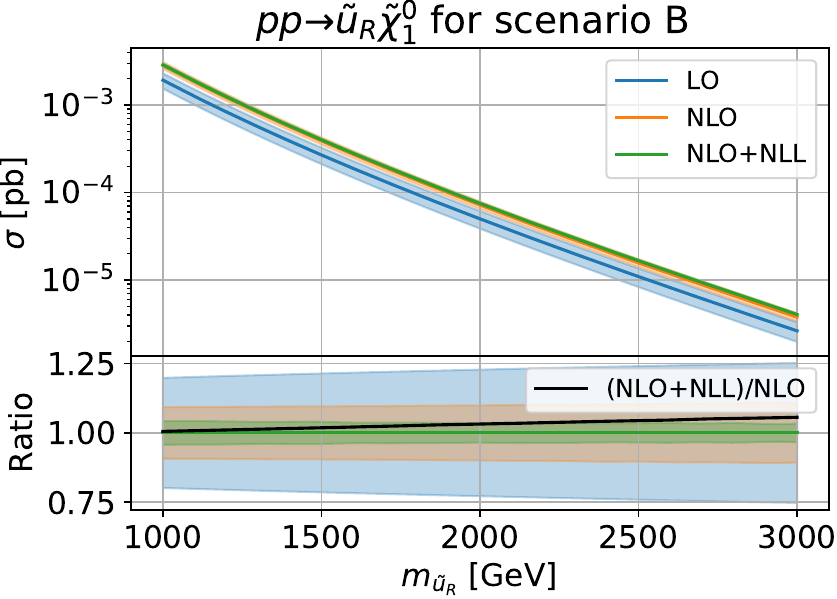}
	\end{subfigure}
	\newline
	\begin{subfigure}[b]{0.49\textwidth}
	    \centering
	     \includegraphics[width=\textwidth]{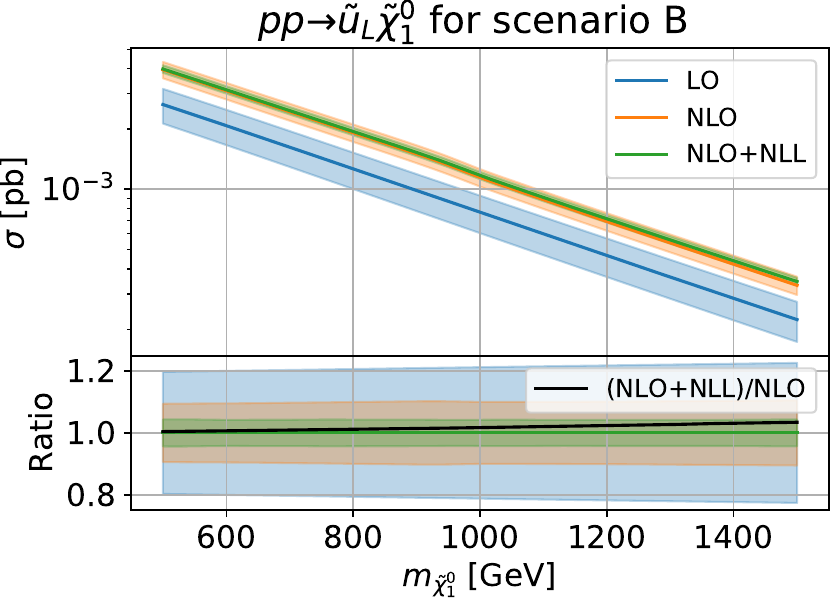}
	\end{subfigure}
	\hfill
	\begin{subfigure}[b]{0.49\textwidth}
	    \centering
	     \includegraphics[width=\textwidth]{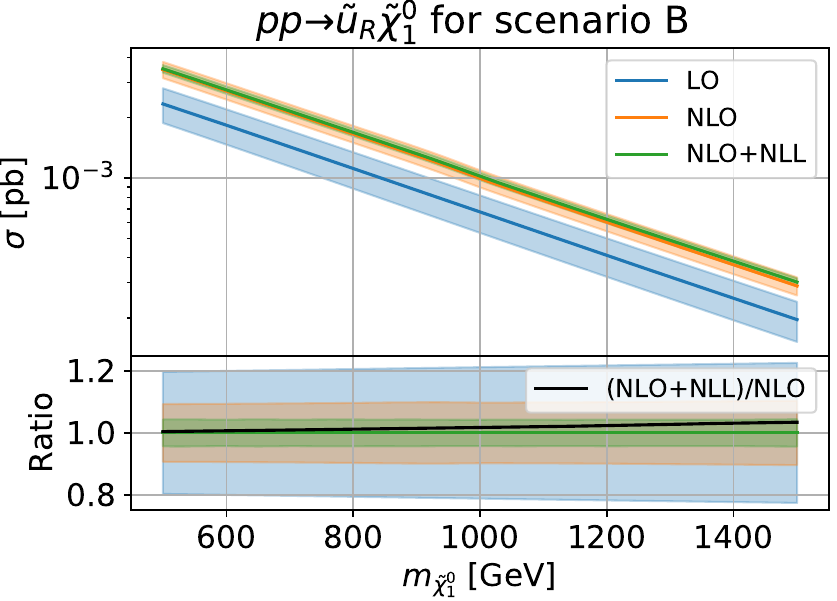}
	\end{subfigure}
	\caption{Total cross sections for the processes $pp \to \upsquark_{L,R}\gaugino^0_1$ and their relative scale uncertainties (top panels) as well as (NLO+NLL)/NLO $K$-factors (bottom panels) in scenario B. In the first row, we vary the squark mass $m_{\upsquark_{L,R}}$, keeping a fixed distance between the left- and right-handed squark masses $m_{\upsquark_L}-m_{\upsquark_R} = \SI{100}{GeV}$. In the second row, we vary the electroweakino mass $m_{\gaugino}$. The other parameters defining scenario B are not modified. The LHC energy is $\sqrt S = \SI{13}{TeV}$, and we use MSHT20 PDFs.}
	\label{fig:mass}
\end{figure}

The dependence of the total cross section for associated squark-electroweakino production on the masses of the produced particles is important to estimate the sensitivity of Run 3 at the LHC to this process. A precise quantitative statement would of course require a detailed signal and background analysis, which is beyond the scope of this work. Therefore, we show in Fig. \ref{fig:mass}
the total cross sections and resulting relative scale uncertainties for scenario B as a function of the SUSY particle masses, both for $\tilde u_L\tilde\chi_1^0$ (left) and $\tilde u_R\tilde\chi_1^0$ (right) production. As expected, the cross sections fall steeply with either mass. Our predictions indicate that an integrated luminosity of \SI{350}{fb^{-1}} at $\sqrt{S} = \SI{13}{TeV}$ from the LHC Run 3~\cite{hllhc} will lead to the production of hundreds of squark-electroweakino events for a neutralino mass of \SI{0.5}{TeV} and squark masses ranging up to \SI{2}{TeV}.

In the lower panels of the plots, we observe an improvement in the precision of the predictions over the whole mass range. Resummation effects reduce the scale dependence from \SI{\pm 10}{\percent} at NLO to below \SI{\pm 5}{\percent} at NLO+NLL. The black curves in the lower insets of the figures represent the ratio of  the NLO+NLL predictions to the NLO ones and  demonstrate the increasing impact of resummation with rising mass values. As in the previous sections, this demonstrates once more that resummation effects are larger near the hadronic threshold. While the central cross section values are enlarged by \SI{50}{\percent} when adding NLO corrections to the LO rates, the additional increase from NLL resummation reaches only about \SI{6}{\percent} for the mass ranges observable at the LHC in the near future.

\section{Conclusion}
\label{sec:concl}

We have presented in this paper a threshold resummation calculation at the NLO+NLL accuracy for the associated production of a squark and an electroweakino at the LHC. This process, like the associated production of a gluino and an electroweakino, has the potential to become important in the near future, if squarks and gluinos turn out to be too heavy to be produced in pairs. The semi-strong production of one electroweak and one strongly-interacting superpartner indeed offers cross sections of intermediate size and a larger available phase space thanks to the possibility of having a lighter electroweakino in the final state.

Our investigations required the calculation of the full NLO corrections to the LO rate, as well as of the associated process-dependent soft anomalous dimension and hard matching coefficients. By matching fixed-order and resummed predictions, we consistently combined the resummation of large logarithms appearing close to threshold at NLL with NLO results. NLL resummation has been found to increase the NLO cross sections for central scale choices by up to \SI{6}{\percent} for squark masses expected to be in the reach of the LHC Run 3. In addition, the resummation procedure allowed for the stabilisation of the predictions relative to the scale dependence, the scale uncertainties being reduced to below \SI{5}{\percent} in the explored mass regime. Our calculation has been included in version 3.1.0 of the public code \textsc{Resummino}~\cite{resumminourl}. It would be interesting to compare in detail our resummation results to those obtained by matching NLO calculations to parton showers \cite{1907.04898,2110.04211}, as it has been done in dedicated publications, {\it e.g.} for $Z'$ boson \cite{Fuks:2007gk} and electroweakino pair production \cite{Fuks:2012qx}.

\acknowledgments
We thank M. Sunder for his collaboration in the early stages of this project. The work of JF has been supported by STFC under the consolidated grant ST/T000988/1. The work of MK and AN has been supported by the BMBF under contract 05P21PMCAA and by the DFG through the Research Training Network 2149 “Strong and Weak Interactions - from Hadrons to Dark Matter”.

\paragraph{Open Access.} 
This article is distributed under the terms of the Creative Commons Attribution License (CC-BY-4.0), which permits any use, distribution and reproduction in any medium, provided the original authors(s) and source are credited.

\appendix
\section{Soft anomalous dimension}
\label{app:soft}
We compute the soft anomalous dimension $\bar{\Gamma}$ for squark-electroweakino production
\begin{equation}
 \bar{\Gamma}_{ab\rightarrow ij} = \Gamma_{ab\rightarrow ij} - \Gamma^{\rm DY}_{ab}
\end{equation}
with
\begin{equation}
 \Gamma_{ab\rightarrow ij} = -\sum_{kl} C^{kl} \lim_{\epsilon\rightarrow 0} \epsilon~\omega^{kl}\,.
\end{equation}
In this expression, the sum goes over eikonal lines, $C^{kl}$ are colour factors, which depend on each specific diagram, $\omega^{kl}$ are integrals over the corresponding kinematic quantities and $\Gamma^{\rm DY}$ includes the contribution from the self-energies of the two incoming Wilson lines. The latter is given by
\begin{equation}
 \Gamma^{\rm DY}_{ab} = \frac{\alpha_s}{2\pi} \sum_{k=\{a,b\}} C_k \left[1 - \log\left(\frac{(v_k \cdot n)^2}{|n|^2}\right) - \log(2) - i\pi\right]\delta_{IJ},
\end{equation}
where the sum goes over the initial-state particles and the colour factor $C_k$ is equal to $C_F$ for an incoming (anti)quark and to $C_A$ for an incoming gluon. The four-momentum $n^\mu$ is an axial gauge vector fulfilling $|n|^2 = -n^2 - i\epsilon$~\cite{Kidonakis:1997gm}. The factor $\delta_{IJ}$ indicates that this contribution has to be subtracted from the diagonal elements in the colour basis of the process. In the process considered in this work, the soft anomalous dimension is a scalar quantity, since we are dealing with a colour basis containing a single element due to the presence of a single coloured final-state particle.

We can write a general expression for the integrals $\omega^{kl}$
\begin{equation}
\begin{split}
 \omega^{kl} = g_s \int & \frac{\dd^D q}{(2\pi)^D} \frac{-i}{q^2 + i\epsilon} \bigg[ \frac{\Delta_k \Delta_l v_k \cdot v_l}{(\delta_k v_k \cdot q + i\epsilon)(\delta_k v_k \cdot q + i\epsilon)} \\
 & - \frac{ \delta_l \Delta_l\Delta_k v_k \cdot n}{\delta_k v_k \cdot q +i\epsilon}\frac{P}{(n \cdot q)} - \frac{\delta_k \Delta_k\Delta_l v_l \cdot n}{\delta_l v_l \cdot q +i\epsilon}\frac{P}{(n \cdot q)} + n^2 \delta_k \Delta_k \delta_l \Delta_l \frac{P}{(n \cdot q)^2} \bigg]\,,
\end{split}
\end{equation}
where we used the eikonal Feynman rules found in Ref.~\cite{rothering:phd}. The above integral depends on the principal value operator $P$, that is calculated as
\begin{equation}
 \frac{P}{(n \cdot q)^\beta} = \frac{1}{2}\left( \frac{1}{(n \cdot q + i\epsilon)^\beta} + (-1)^\beta \frac{1}{(-n \cdot q + i\epsilon)^\beta} \right)\,,
\end{equation}
on $v_i^\mu = p_i^\mu \sqrt{2/s}$, and on the process-dependent signs $\Delta$ and $\delta$.

\begin{figure}
\begin{center}
\includegraphics[width=0.6\textwidth]{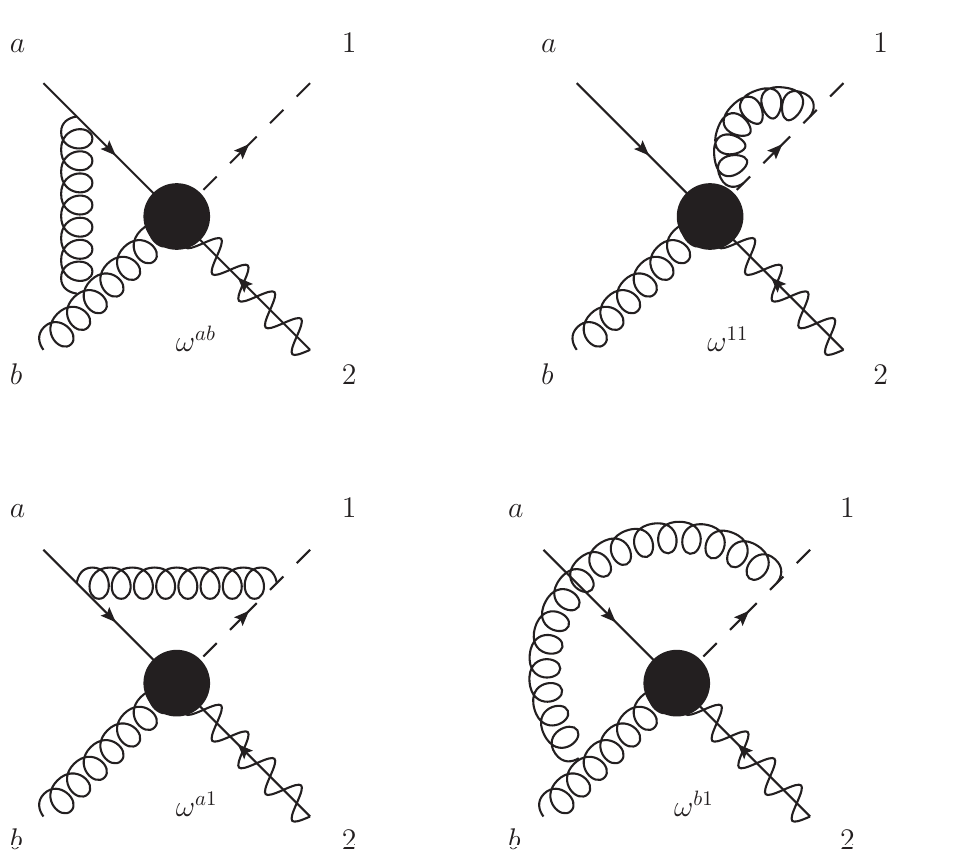}
\caption{Eikonal diagrams for the soft anomalous dimension in squark-electroweakino production.}
\label{fig:gaugino_squark}
\end{center}
\end{figure}

We use the solutions of Ref.~\cite{theeuwe:phd} for the integrals associated with the diagrams in \cref{fig:gaugino_squark}. We obtain similar expressions as in Ref.~\cite{1604.01023}, apart from the combined signs $S_{kl}=\Delta_k\Delta_l \delta_k \delta_l$ that are given by $S_{ab} = -1$, $S_{a1}=+1$, $ S_{b1}=+1$ and $S_{11}=-1$. To be specific, we obtain
\begin{align}
 \omega^{ab} &= S_{ab}\frac{\alpha_s}{\epsilon\pi} \left[ -\log\left(\frac{v_a \cdot v_b}{2}\right) + \frac{1}{2}\log\left(\frac{(v_a \cdot n)^2}{|n|^2}\frac{(v_b \cdot n)^2}{|n|^2}\right) + i\pi - 1 \right]
 \,,\\
 \omega^{a1} &= S_{a1}\frac{\alpha_s}{\epsilon\pi} \left[ -\frac{1}{2}\log\left(\frac{(v_a \cdot v_1)^2 s}{2 m_1^2}\right) + L_1 + \frac{1}{2}\log\left(\frac{(v_a \cdot n)^2}{|n|^2}\right) - 1 \right]
 \,,\\
 \omega^{b1} &= S_{b1}\frac{\alpha_s}{\epsilon\pi} \left[ -\frac{1}{2}\log\left(\frac{(v_b \cdot v_1)^2 s}{2 m_1^2}\right) + L_1 + \frac{1}{2}\log\left(\frac{(v_b \cdot n)^2}{|n|^2}\right) - 1 \right]
 \,,\\
 \omega^{11} &= S_{11}\frac{\alpha_s}{\epsilon\pi} \left[ 2 L_1 - 2 \right]
 \,.
\end{align}

For squark-electroweakino production, the final-state massive particle is a squark, so that the kinematic quantities read
\begin{align}
 v_a \cdot v_b \!=\! \frac{2 p_a \cdot p_b}{s} \!=\! 1,\ \
 v_a \cdot v_1 \!=\! \frac{2 p_a \cdot p_1}{s} \!=\! \frac{m_{\tilde{q}}^2 - t}{s}
 \ \ \text{and}\ \
 v_b \cdot v_1 \!=\! \frac{2 p_b \cdot p_1}{s} \!=\! \frac{m_{\tilde{q}}^2 - u}{s}\,,
\end{align}
where $t$ and $u$ are the usual Mandelstam variables.
The remaining quantity $L_1$ simplifies to one in the limit $(v_1 \cdot n)^2 \rightarrow 2 m_{\tilde{q}}^2 n^2 / s$.

The relevant colour factors $C^{kl}$ are
\begin{equation}
\begin{aligned}
 C^{ab} &= \frac{{\rm Tr[T^i T^{i'} T^j]} (-i) f^{i'ij}}{{\rm Tr[T^i T^i]}} = -\frac{C_A}{2}\,, \quad &&
 C^{11} = \frac{{\rm Tr[T^i T^j T^j T^i]}}{{\rm Tr[T^i T^i]}} = C_F\,,  \\
 C^{a1} &= \frac{{\rm Tr[T^i T^j T^i T^j]}}{{\rm Tr[T^i T^i]}} = C_F - \frac{C_A}{2}\,, \quad && 
 C^{b1} = \frac{{\rm Tr[T^i T^j T^{i'}]} (-i) f^{i'ij}}{{\rm Tr[T^i T^i]}} = \frac{C_A}{2}  \,,
\end{aligned}
\end{equation}
where we have used the well-known relation $T^a T^a = C_F {\bf 1}$.

With these ingredients we can calculate $\Gamma_{ab\rightarrow ij}$
\begin{alignat}{4}
 \Gamma_{ab\rightarrow ij} ={}& \epsilon&&\bigg[ &&-&&C_F (\omega^{a1} + \omega^{11}) + \frac{C_A}{2}(\omega^{ab} + \omega^{a1} - \omega^{b1}) \bigg]  
  \\\nonumber={}& \frac{\alpha_s}{2\pi}&&\bigg[ &&C_F &&\bigg( -\log\left(\frac{(v_a \cdot n)^2}{|n|^2}\right) + \log\left(\frac{(v_a \cdot v_1)^2 s}{2 m_{\tilde{q}}^2}\right) + 2 L_1 - 2 \bigg) 
  \\\nonumber& &&+&&C_A &&\bigg( -\frac{1}{2}\log\left(\frac{(v_a \cdot v_1)^2 s}{2 m_{\tilde{q}}^2}\right) + \frac{1}{2}\log\left(\frac{(v_b \cdot v_1)^2 s}{2 m_{\tilde{q}}^2}\right) 
  \\&&&&&&&- \log\left(\frac{(v_b \cdot n)^2}{|n|^2}\right) + \log\left(\frac{v_a \cdot v_b}{2}\right) + 1 - i\pi \bigg) \bigg]
  \,,
\end{alignat}
while the contribution from the initial-state self energies is
\begin{equation}
\begin{split}
 \Gamma^{\rm DY}_{ab} = \frac{\alpha_s}{2\pi} \bigg[ &C_F \left(1 - \log\left(\frac{(v_a \cdot n)^2}{|n|^2}\right) - \log(2) - i\pi\right) 
 \\+&C_A \left(1 - \log\left(\frac{(v_b \cdot n)^2}{|n|^2}\right)- \log(2) - i\pi\right)\bigg]
 \,.
\end{split}
\end{equation}
We observe the cancellation of the gauge dependent terms. This gives as the final result for the soft anomalous dimension
\begin{align}\nonumber 
 \bar{\Gamma}_{ab\rightarrow ij} ={} \frac{\alpha_s}{2\pi}\bigg\{ &C_F \left[ \log\left(\frac{(v_a \cdot v_1)^2 s}{2 m_{\tilde{q}}^2}\right) + 2 L_1 - 3 + \log(2) +i\pi \right] +
  \\+{} &C_A \left[ -\frac{1}{2}\log\left(\frac{(v_a \cdot v_1)^2 s}{2 m_{\tilde{q}}^2}\right) + \frac{1}{2}\log\left(\frac{(v_b \cdot v_1)^2 s}{2 m_{\tilde{q}}^2}\right) + \log\left(\frac{v_a \cdot v_b }{2}\right) + \log(2) \right] \bigg\} 
 \nonumber\\ ={} \frac{\alpha_s}{2\pi}\bigg\{ &C_F \left[ 2\log\left(\frac{m_{\tilde{q}}^2 - t}{\sqrt{s} m_{\tilde{q}}}\right) - 1 + i\pi \right] + C_A  \log\left(\frac{m_{\tilde{q}}^2 - u}{m_{\tilde{q}}^2 - t}\right) \bigg\}
 \,.
\end{align}
This result is related to the one for $tW$ production \cite{Kidonakis:2006bu} and in the massless limit also to the one for the QCD Compton process \cite{Laenen:1998qw}. It agrees in particular with eq.\ (3.8) of Ref.\ \cite{Kidonakis:2006bu} after subtraction of the Drell-Yan terms.
   
\section{Lists of total cross sections}
\label{app:table}

In \cref{tab:squark} and \cref{tab:gaugino}, we show the total cross sections for the associated production of an up-type squark and the lightest neutralino at NLO and NLO+NLL accuracy for an LHC centre-of-mass energy of $\sqrt{S} = \SI{13}{TeV}$ in scenario B. The results are listed both as a function of the squark mass and of the neutralino mass and correspond to the plots in \cref{fig:mass}. We give the scale and PDF uncertainty for the resummed result for the three PDFs we already investigated in \cref{fig:pdfs}, namely MSHT20NLO118, CT18NLO and NNPDF40.

\renewcommand{\arraystretch}{1.55}
\begin{table}
\begin{center}
	\begin{tabular}{ |c|c|c|c|c| } 
		\hline
		$m_{\upsquark_L}$ [GeV] &  NLO$_{-\text{scale}-\text{PDF}}^{+\text{scale}+\text{PDF}}$ [fb] &\multicolumn{3}{c|}{NLO+NLL$_{-\text{scale}-\text{PDF}}^{+\text{scale}+\text{PDF}}$ [fb]}  \\
		\hline
		&  MSHT20NLO118 & MSHT20NLO118 & CT18NLO & NNPDF40 \\
		\hline
$1000$ 	&	 $3.301^{+9.5\%+2.4\%}_{-9.5\%-2.0\%}$ 		&	 $3.318^{+4.7\%+2.6\%}_{-3.9\%-1.8\%}$ 				&	 $3.353^{+4.7\%+5.2\%}_{-3.9\%-3.4\%}$ 				&	 $3.234^{+4.7\%+2.4\%}_{-3.9\%-2.4\%}$ 					\\
$1100$ 	&	 $2.148^{+9.6\%+2.7\%}_{-9.5\%-2.1\%}$ 		&	 $2.166^{+4.5\%+3.1\%}_{-3.8\%-1.6\%}$ 				&	 $2.188^{+4.6\%+5.5\%}_{-3.8\%-3.6\%}$ 				&	 $2.105^{+4.6\%+2.4\%}_{-3.8\%-2.4\%}$ 					\\
$1200$ 	&	 $1.426^{+9.6\%+2.8\%}_{-9.6\%-2.1\%}$ 		&	 $1.442^{+4.4\%+3.3\%}_{-3.8\%-1.6\%}$ 				&	 $1.456^{+4.5\%+6.0\%}_{-3.7\%-3.7\%}$ 				&	 $1.398^{+4.5\%+2.5\%}_{-3.7\%-2.5\%}$ 					\\
$1300$ 	&	 $0.963^{+9.7\%+3.0\%}_{-9.7\%-2.3\%}$ 		&	 $0.976^{+4.3\%+3.5\%}_{-3.7\%-1.7\%}$ 				&	 $0.986^{+4.3\%+6.1\%}_{-3.7\%-4.0\%}$ 				&	 $0.944^{+4.4\%+2.6\%}_{-3.7\%-2.6\%}$ 					\\
$1400$ 	&	 $0.660^{+9.7\%+3.1\%}_{-9.9\%-2.4\%}$ 		&	 $0.671^{+4.2\%+3.4\%}_{-3.7\%-2.1\%}$ 				&	 $0.678^{+4.3\%+6.5\%}_{-3.6\%-4.2\%}$ 				&	 $0.647^{+4.4\%+2.7\%}_{-3.6\%-2.7\%}$ 					\\
$1500$ 	&	 $0.458^{+9.8\%+3.4\%}_{-10.0\%-2.5\%}$ 	&	 $0.467^{+4.2\%+3.7\%}_{-3.6\%-2.1\%}$ 				&	 $0.472^{+4.2\%+6.9\%}_{-3.6\%-4.4\%}$ 				&	 $0.450^{+4.3\%+2.7\%}_{-3.6\%-2.7\%}$ 					\\
$1600$ 	&	 $0.322^{+9.9\%+3.6\%}_{-10.1\%-2.6\%}$ 	&	 $0.329^{+4.1\%+4.0\%}_{-3.5\%-2.2\%}$ 				&	 $0.332^{+4.1\%+7.3\%}_{-3.5\%-4.5\%}$ 				&	 $0.316^{+4.2\%+2.8\%}_{-3.5\%-2.8\%}$ 					\\
$1700$ 	&	 $0.229^{+10.0\%+3.6\%}_{-10.2\%-2.9\%}$ 	&	 $0.234^{+4.0\%+3.9\%}_{-3.5\%-2.3\%}$					&	 $0.237^{+4.1\%+7.7\%}_{-3.5\%-4.7\%}$					&	 $0.224^{+4.2\%+2.9\%}_{-3.4\%-2.9\%}$					 \\
$1800$ 	&	 $0.164^{+10.2\%+4.2\%}_{-10.4\%-2.7\%}$ 	&	 $0.168^{+4.0\%+4.6\%}_{-3.5\%-2.2\%}$					&	 $0.170^{+4.0\%+8.0\%}_{-3.4\%-4.9\%}$					&	 $0.161^{+4.1\%+3.0\%}_{-3.4\%-3.0\%}$					 \\
$1900$ 	&	 $0.118^{+10.2\%+4.1\%}_{-10.5\%-3.2\%}$ 	&	 $0.122^{+3.9\%+4.5\%}_{-3.5\%-2.6\%}$					&	 $0.123^{+3.9\%+8.0\%}_{-3.4\%-5.6\%}$					&	 $0.116^{+4.1\%+3.1\%}_{-3.4\%-3.1\%}$					 \\
$2000$ 	&	 $0.086^{+10.4\%+4.4\%}_{-10.6\%-3.4\%}$ 	&	 $0.089^{+3.9\%+4.8\%}_{-3.4\%-2.8\%}$					&	 $0.090^{+3.9\%+8.6\%}_{-3.3\%-5.5\%}$					&	 $0.084^{+4.0\%+3.2\%}_{-3.3\%-3.2\%}$					 \\
$2100$ 	&	 $0.063^{+10.5\%+4.7\%}_{-10.7\%-3.3\%}$ 	&	 $0.065^{+3.8\%+4.9\%}_{-3.4\%-2.9\%}$					&	 $0.066^{+3.8\%+8.5\%}_{-3.4\%-6.3\%}$					&	 $0.062^{+4.0\%+3.4\%}_{-3.3\%-3.4\%}$					 \\
$2200$ 	&	 $0.046^{+10.7\%+5.2\%}_{-10.9\%-3.4\%}$ 	&	 $0.048^{+3.8\%+5.3\%}_{-3.3\%-3.1\%}$					&	 $0.049^{+3.8\%+9.4\%}_{-3.3\%-6.1\%}$					&	 $0.046^{+4.0\%+3.5\%}_{-3.2\%-3.5\%}$					 \\
$2300$ 	&	 $0.034^{+10.9\%+5.3\%}_{-11.1\%-3.7\%}$ 	&	 $0.036^{+3.7\%+5.3\%}_{-3.3\%-3.4\%}$					&	 $0.036^{+3.8\%+9.7\%}_{-3.3\%-6.4\%}$					&	 $0.034^{+3.9\%+3.6\%}_{-3.2\%-3.6\%}$					 \\
$2400$ 	&	 $0.026^{+10.9\%+5.8\%}_{-11.2\%-3.7\%}$ 	&	 $0.027^{+3.7\%+5.9\%}_{-3.2\%-3.3\%}$				 	&	 $0.027^{+3.7\%+10.2\%}_{-3.3\%-6.6\%}$			&	 $0.025^{+3.8\%+3.7\%}_{-3.2\%-3.7\%}$					 \\
$2500$ 	&	 $0.019^{+11.2\%+6.3\%}_{-11.3\%-3.6\%}$ 	&	 $0.020^{+3.7\%+6.4\%}_{-3.2\%-3.3\%}$				 	&	 $0.020^{+3.6\%+10.5\%}_{-3.3\%-7.0\%}$			&	 $0.019^{+3.7\%+3.8\%}_{-3.3\%-3.8\%}$					 \\
$2600$ 	&	 $0.014^{+11.3\%+6.3\%}_{-11.4\%-4.1\%}$ 	&	 $0.015^{+3.5\%+6.2\%}_{-3.3\%-3.9\%}$				 	&	 $0.015^{+3.6\%+10.9\%}_{-3.3\%-7.2\%}$			&	 $0.014^{+3.8\%+4.0\%}_{-3.1\%-4.0\%}$					 \\
$2700$ 	&	 $0.011^{+11.5\%+7.1\%}_{-11.6\%-3.9\%}$ 	&	 $0.011^{+3.5\%+7.0\%}_{-3.2\%-3.7\%}$				 	&	 $0.011^{+3.5\%+11.3\%}_{-3.2\%-7.6\%}$			&	 $0.011^{+3.7\%+4.1\%}_{-3.1\%-4.1\%}$					 \\
$2800$ 	&	 $0.008^{+11.7\%+7.6\%}_{-11.7\%-4.1\%}$ 	&	 $0.009^{+3.5\%+7.5\%}_{-3.2\%-3.9\%}$				 	&	 $0.009^{+3.4\%+11.9\%}_{-3.2\%-7.8\%}$			&	 $0.008^{+3.6\%+4.2\%}_{-3.2\%-4.2\%}$					 \\
$2900$ 	&	 $0.006^{+11.9\%+8.2\%}_{-11.8\%-4.1\%}$ 	&	 $0.006^{+3.4\%+8.0\%}_{-3.1\%-4.0\%}$				 	&	 $0.007^{+3.3\%+12.4\%}_{-3.2\%-7.9\%}$			&	 $0.006^{+3.5\%+4.4\%}_{-3.2\%-4.4\%}$					 \\
$3000$ 	&	 $0.005^{+11.8\%+8.5\%}_{-11.9\%-4.2\%}$ 	&	 $0.005^{+3.4\%+8.3\%}_{-3.0\%-4.0\%}$				 	&	 $0.005^{+3.3\%+13.0\%}_{-3.2\%-8.0\%}$			&	 $0.005^{+3.5\%+4.5\%}_{-3.1\%-4.5\%}$
	 \\\hline
	\end{tabular}
\end{center}
\caption{
	Total cross sections for $pp \to \upsquark_L\gaugino_1^0$ at $\sqrt{S} = \SI{13}{TeV}$ at NLO and NLO+NLL for various squark masses, with a fixed mass difference between left- and right-handed squarks of $m_{\upsquark_L}-m_{\upsquark_R} = \SI{100}{GeV}$, and PDFs. The remaining parameters are fixed to scenario B.
}
\label{tab:squark}
\end{table}

\renewcommand{\arraystretch}{1.6}\begin{table}
\begin{center}
	\begin{tabular}{ |c|c|c|c|c| } 
		\hline
		$m_{\gaugino^0_1}$ [GeV]&  NLO$_{-\text{scale}-\text{PDF}}^{+\text{scale}+\text{PDF}}$ [fb] &\multicolumn{3}{c|}{NLO+NLL$_{-\text{scale}-\text{PDF}}^{+\text{scale}+\text{PDF}}$ [fb]} \\
		\hline
		&  MSHT20NLO118 & MSHT20NLO118 & CT18NLO & NNPDF40 \\
		\hline
$500$ 	&	 $3.950^{+9.5\%+2.3\%}_{-9.4\%-2.3\%}$ 		&	 $3.966^{+4.7\%+2.4\%}_{-3.9\%-2.1\%}$ 			&	 $4.007^{+4.8\%+4.8\%}_{-3.9\%-3.6\%}$ 						&	 $3.867^{+4.8\%+2.3\%}_{-3.9\%-2.3\%}$ 				\\
$550$ 	&	 $3.502^{+9.6\%+2.4\%}_{-9.5\%-2.2\%}$ 		&	 $3.521^{+4.7\%+2.6\%}_{-3.8\%-2.0\%}$ 			&	 $3.556^{+4.8\%+5.0\%}_{-3.8\%-3.6\%}$ 						&	 $3.429^{+4.8\%+2.4\%}_{-3.8\%-2.4\%}$ 				\\
$600$ 	&	 $3.102^{+9.6\%+2.4\%}_{-9.6\%-2.2\%}$ 		&	 $3.123^{+4.7\%+2.5\%}_{-3.8\%-2.0\%}$ 			&	 $3.154^{+4.8\%+5.1\%}_{-3.8\%-3.6\%}$ 						&	 $3.039^{+4.8\%+2.4\%}_{-3.8\%-2.4\%}$ 				\\
$650$ 	&	 $2.748^{+9.7\%+2.6\%}_{-9.6\%-2.1\%}$ 		&	 $2.769^{+4.7\%+2.7\%}_{-3.8\%-2.0\%}$ 			&	 $2.797^{+4.7\%+5.4\%}_{-3.8\%-3.7\%}$ 						&	 $2.692^{+4.8\%+2.5\%}_{-3.8\%-2.5\%}$ 				\\
$700$ 	&	 $2.434^{+9.8\%+2.7\%}_{-9.7\%-2.1\%}$ 		&	 $2.456^{+4.7\%+2.9\%}_{-3.8\%-1.9\%}$ 			&	 $2.481^{+4.7\%+5.5\%}_{-3.8\%-3.8\%}$ 						&	 $2.385^{+4.8\%+2.5\%}_{-3.8\%-2.5\%}$ 				\\
$750$ 	&	 $2.158^{+9.9\%+2.7\%}_{-9.8\%-2.2\%}$ 		&	 $2.180^{+4.6\%+2.8\%}_{-3.9\%-2.0\%}$ 			&	 $2.202^{+4.7\%+5.6\%}_{-3.8\%-3.8\%}$ 						&	 $2.115^{+4.7\%+2.5\%}_{-3.9\%-2.5\%}$ 				\\
$800$ 	&	 $1.913^{+10.1\%+2.6\%}_{-9.9\%-2.4\%}$ 	&	 $1.936^{+4.6\%+2.8\%}_{-3.9\%-2.2\%}$ 			&	 $1.955^{+4.6\%+5.9\%}_{-3.9\%-3.8\%}$ 						&	 $1.876^{+4.6\%+2.5\%}_{-3.9\%-2.5\%}$ 				\\
$850$ 	&	 $1.697^{+10.2\%+3.0\%}_{-10.0\%-2.1\%}$ 	&	 $1.718^{+4.5\%+3.1\%}_{-3.9\%-1.9\%}$				&	 $1.736^{+4.6\%+6.2\%}_{-3.9\%-3.6\%}$							&	 $1.664^{+4.6\%+2.6\%}_{-3.9\%-2.6\%}$				 \\
$900$ 	&	 $1.502^{+10.3\%+3.0\%}_{-10.1\%-2.2\%}$ 	&	 $1.523^{+4.5\%+3.0\%}_{-3.9\%-2.0\%}$				&	 $1.538^{+4.5\%+6.4\%}_{-3.9\%-3.8\%}$							&	 $1.473^{+4.6\%+2.6\%}_{-3.9\%-2.6\%}$				 \\
$950$ 	&	 $1.319^{+10.3\%+3.3\%}_{-10.1\%-2.1\%}$ 	&	 $1.340^{+4.6\%+3.4\%}_{-3.8\%-1.9\%}$				&	 $1.353^{+4.6\%+6.6\%}_{-3.8\%-3.8\%}$							&	 $1.295^{+4.6\%+2.7\%}_{-3.8\%-2.7\%}$				 \\
$1000$ 	&	 $1.146^{+10.0\%+3.2\%}_{-10.0\%-2.4\%}$ 	&	 $1.166^{+4.7\%+3.3\%}_{-3.7\%-2.2\%}$		 	&	 $1.178^{+4.8\%+6.8\%}_{-3.7\%-4.0\%}$					 	&	 $1.125^{+4.8\%+2.7\%}_{-3.7\%-2.7\%}$							\\
$1050$ 	&	 $1.007^{+10.0\%+3.4\%}_{-10.0\%-2.4\%}$ 	&	 $1.025^{+4.9\%+3.7\%}_{-3.6\%-2.0\%}$			&	 $1.036^{+4.9\%+7.2\%}_{-3.6\%-3.9\%}$ 					 	&	 $0.988^{+5.0\%+2.8\%}_{-3.6\%-2.8\%}$							\\
$1100$ 	&	 $0.887^{+10.0\%+3.6\%}_{-10.1\%-2.4\%}$ 	&	 $0.904^{+4.9\%+4.0\%}_{-3.5\%-1.9\%}$			&	 $0.914^{+5.0\%+7.7\%}_{-3.5\%-4.0\%}$ 						&	 $0.870^{+5.1\%+2.8\%}_{-3.5\%-2.8\%}$ 							\\
$1150$ 	&	 $0.782^{+10.0\%+3.8\%}_{-10.1\%-2.4\%}$ 	&	 $0.799^{+5.0\%+4.1\%}_{-3.5\%-2.0\%}$			&	 $0.808^{+5.1\%+7.9\%}_{-3.5\%-4.0\%}$ 					 	&	 $0.768^{+5.2\%+2.9\%}_{-3.5\%-2.9\%}$							\\
$1200$ 	&	 $0.691^{+10.0\%+4.0\%}_{-10.2\%-2.5\%}$ 	&	 $0.706^{+5.1\%+4.3\%}_{-3.5\%-2.0\%}$		 	&	 $0.715^{+5.2\%+8.2\%}_{-3.5\%-4.2\%}$					 	&	 $0.679^{+5.2\%+2.9\%}_{-3.5\%-2.9\%}$							\\
$1250$ 	&	 $0.610^{+10.1\%+4.1\%}_{-10.2\%-2.5\%}$ 	&	 $0.625^{+5.2\%+4.6\%}_{-3.5\%-1.8\%}$		 	&	 $0.633^{+5.2\%+8.3\%}_{-3.5\%-4.3\%}$					 	&	 $0.600^{+5.3\%+2.9\%}_{-3.5\%-2.9\%}$							\\
$1300$ 	&	 $0.540^{+10.1\%+4.3\%}_{-10.3\%-2.5\%}$ 	&	 $0.554^{+5.3\%+5.2\%}_{-3.5\%-1.4\%}$		 	&	 $0.560^{+5.3\%+8.4\%}_{-3.5\%-4.6\%}$					 	&	 $0.531^{+5.4\%+3.0\%}_{-3.5\%-3.0\%}$							\\
$1350$ 	&	 $0.478^{+10.2\%+4.0\%}_{-10.3\%-3.1\%}$ 	&	 $0.491^{+5.4\%+4.9\%}_{-3.5\%-2.1\%}$		 	&	 $0.497^{+5.4\%+8.4\%}_{-3.6\%-4.9\%}$					 	&	 $0.470^{+5.5\%+3.0\%}_{-3.6\%-3.0\%}$							\\
$1400$ 	&	 $0.423^{+10.3\%+4.1\%}_{-10.4\%-3.2\%}$ 	&	 $0.435^{+5.4\%+5.2\%}_{-3.6\%-2.0\%}$		 	&	 $0.440^{+5.5\%+8.4\%}_{-3.6\%-5.3\%}$					 	&	 $0.416^{+5.6\%+3.1\%}_{-3.7\%-3.1\%}$							\\
$1450$ 	&	 $0.374^{+10.3\%+4.2\%}_{-10.5\%-3.3\%}$ 	&	 $0.386^{+5.5\%+5.3\%}_{-3.6\%-2.1\%}$		 	&	 $0.391^{+5.5\%+8.3\%}_{-3.7\%-5.6\%}$					 	&	 $0.369^{+5.6\%+3.1\%}_{-3.7\%-3.1\%}$							\\
$1500$ 	&	 $0.332^{+10.4\%+4.3\%}_{-10.6\%-3.5\%}$ 	&	 $0.342^{+5.6\%+5.5\%}_{-3.7\%-2.2\%}$		 	&	 $0.347^{+5.6\%+8.7\%}_{-3.7\%-5.6\%}$					 	&	 $0.327^{+5.7\%+3.2\%}_{-3.8\%-3.2\%}$
	 \\\hline
	\end{tabular}
\end{center}
\caption{
	Total cross sections for $pp \to \upsquark_L\gaugino_1^0$ at $\sqrt{S} = \SI{13}{TeV}$ at NLO and NLO+NLL for various neutralino masses and PDFs. The remaining parameters are fixed to scenario B.
}
\label{tab:gaugino}

\end{table}

\bibliography{References}

\providecommand{\href}[2]{#2}\begingroup\raggedright\begin{thebibliography}{100}

\bibitem{Wess:1974tw}
J.~Wess and B.~Zumino, {\it {Supergauge Transformations in Four-Dimensions}},
  {\em Nucl. Phys. B} {\bf 70} (1974) 39--50.

\bibitem{Salam:1974yz}
A.~Salam and J.~A. Strathdee, {\it {Supergauge Transformations}},  {\em Nucl.
  Phys. B} {\bf 76} (1974) 477--482.

\bibitem{Ferrara:1974pu}
S.~Ferrara and B.~Zumino, {\it {Supergauge Invariant Yang-Mills Theories}},
  {\em Nucl. Phys. B} {\bf 79} (1974) 413.

\bibitem{Sakai:1981gr}
N.~Sakai, {\it {Naturalness in Supersymmetric Guts}},  {\em Z. Phys. C} {\bf
  11} (1981) 153.

\bibitem{Chamseddine:1982jx}
A.~H. Chamseddine, R.~L. Arnowitt, and P.~Nath, {\it {Locally Supersymmetric
  Grand Unification}},  {\em Phys. Rev. Lett.} {\bf 49} (1982) 970.

\bibitem{Ellis:1990wk}
J.~R. Ellis, S.~Kelley, and D.~V. Nanopoulos, {\it {Probing the desert using
  gauge coupling unification}},  {\em Phys. Lett. B} {\bf 260} (1991) 131--137.

\bibitem{Ellis:1983ew}
J.~R. Ellis, J.~S. Hagelin, D.~V. Nanopoulos, K.~A. Olive, and M.~Srednicki,
  {\it {Supersymmetric Relics from the Big Bang}},  {\em Nucl. Phys. B} {\bf
  238} (1984) 453--476.

\bibitem{Jungman:1995df}
G.~Jungman, M.~Kamionkowski, and K.~Griest, {\it {Supersymmetric dark matter}},
   {\em Phys. Rept.} {\bf 267} (1996) 195--373,
  [\href{http://arxiv.org/abs/hep-ph/9506380}{{\tt hep-ph/9506380}}].

\bibitem{Fayet:1976cr}
P.~Fayet and S.~Ferrara, {\it {Supersymmetry}},  {\em Phys. Rept.} {\bf 32}
  (1977) 249--334.

\bibitem{Inoue:1982pi}
K.~Inoue, A.~Kakuto, H.~Komatsu, and S.~Takeshita, {\it {Aspects of Grand
  Unified Models with Softly Broken Supersymmetry}},  {\em Prog. Theor. Phys.}
  {\bf 68} (1982) 927. [Erratum: Prog.Theor.Phys. 70, 330 (1983)].

\bibitem{Nilles:1983ge}
H.~P. Nilles, {\it {Supersymmetry, Supergravity and Particle Physics}},  {\em
  Phys. Rept.} {\bf 110} (1984) 1--162.

\bibitem{Haber:1984rc}
H.~E. Haber and G.~L. Kane, {\it {The Search for Supersymmetry: Probing Physics
  Beyond the Standard Model}},  {\em Phys. Rept.} {\bf 117} (1985) 75--263.

\bibitem{Witten:1981nf}
E.~Witten, {\it {Dynamical Breaking of Supersymmetry}},  {\em Nucl. Phys. B}
  {\bf 188} (1981) 513.

\bibitem{Girardello:1981wz}
L.~Girardello and M.~T. Grisaru, {\it {Soft Breaking of Supersymmetry}},  {\em
  Nucl. Phys. B} {\bf 194} (1982) 65.

\bibitem{Farrar:1978xj}
G.~R. Farrar and P.~Fayet, {\it {Phenomenology of the Production, Decay, and
  Detection of New Hadronic States Associated with Supersymmetry}},  {\em Phys.
  Lett. B} {\bf 76} (1978) 575--579.

\bibitem{hllhc}
I.~Zurbano~Fernandez et~al., {\it {High-Luminosity Large Hadron Collider
  (HL-LHC): Technical design report}},  CERN-2020-010.

\bibitem{CidVidal:2018eel}
X.~Cid~Vidal et~al., {\it {Report from Working Group 3}: {Beyond the Standard
  Model physics at the HL-LHC and HE-LHC}},  {\em CERN Yellow Rep. Monogr.}
  {\bf 7} (2019) 585--865, [\href{http://arxiv.org/abs/1812.07831}{{\tt
  arXiv:1812.07831}}].

\bibitem{Gianotti:2002xx}
F.~Gianotti et~al., {\it {Physics potential and experimental challenges of the
  LHC luminosity upgrade}},  {\em Eur. Phys. J. C} {\bf 39} (2005) 293--333,
  [\href{http://arxiv.org/abs/hep-ph/0204087}{{\tt hep-ph/0204087}}].

\bibitem{1108.1250}
T.~Binoth, D.~Goncalves~Netto, D.~Lopez-Val, K.~Mawatari, T.~Plehn, and
  I.~Wigmore, {\it {Automized Squark-Neutralino Production to Next-to-Leading
  Order}},  {\em Phys. Rev. D} {\bf 84} (2011) 075005,
  [\href{http://arxiv.org/abs/1108.1250}{{\tt arXiv:1108.1250}}].

\bibitem{1907.04898}
S.~Frixione, B.~Fuks, V.~Hirschi, K.~Mawatari, H.-S. Shao, P.~A. Sunder, and
  M.~Zaro, {\it {Automated simulations beyond the Standard Model:
  supersymmetry}},  {\em JHEP} {\bf 12} (2019) 008,
  [\href{http://arxiv.org/abs/1907.04898}{{\tt arXiv:1907.04898}}].

\bibitem{2110.04211}
J.~Baglio, G.~Coniglio, B.~Jager, and M.~Spira, {\it {Next-to-leading-order QCD
  corrections and parton-shower effects for weakino+squark production at the
  LHC}},  {\em JHEP} {\bf 12} (2021) 020,
  [\href{http://arxiv.org/abs/2110.04211}{{\tt arXiv:2110.04211}}].

\bibitem{Sterman:1986aj}
G.~F. Sterman, {\it {Summation of Large Corrections to Short Distance Hadronic
  Cross-Sections}},  {\em Nucl. Phys. B} {\bf 281} (1987) 310--364.

\bibitem{Catani1989}
S.~Catani and L.~Trentadue, {\it {Resummation of the QCD Perturbative Series
  for Hard Processes}},  {\em Nucl. Phys. B} {\bf 327} (1989) 323--352.

\bibitem{Catani1991}
S.~Catani and L.~Trentadue, {\it {Comment on QCD exponentiation at large x}},
  {\em Nucl. Phys. B} {\bf 353} (1991) 183--186.

\bibitem{Kidonakis:1997gm}
N.~Kidonakis and G.~F. Sterman, {\it {Resummation for QCD hard scattering}},
  {\em Nucl. Phys. B} {\bf 505} (1997) 321--348,
  [\href{http://arxiv.org/abs/hep-ph/9705234}{{\tt hep-ph/9705234}}].

\bibitem{hep-ph/9801268}
N.~Kidonakis, G.~Oderda, and G.~F. Sterman, {\it {Threshold resummation for
  dijet cross-sections}},  {\em Nucl. Phys. B} {\bf 525} (1998) 299--332,
  [\href{http://arxiv.org/abs/hep-ph/9801268}{{\tt hep-ph/9801268}}].

\bibitem{Bozzi:2004qq}
G.~Bozzi, B.~Fuks, and M.~Klasen, {\it {Slepton production in polarized hadron
  collisions}},  {\em Phys. Lett. B} {\bf 609} (2005) 339--350,
  [\href{http://arxiv.org/abs/hep-ph/0411318}{{\tt hep-ph/0411318}}].

\bibitem{Bozzi:2006fw}
G.~Bozzi, B.~Fuks, and M.~Klasen, {\it {Transverse-momentum resummation for
  slepton-pair production at the CERN LHC}},  {\em Phys. Rev. D} {\bf 74}
  (2006) 015001, [\href{http://arxiv.org/abs/hep-ph/0603074}{{\tt
  hep-ph/0603074}}].

\bibitem{Bozzi:2007qr}
G.~Bozzi, B.~Fuks, and M.~Klasen, {\it {Threshold Resummation for Slepton-Pair
  Production at Hadron Colliders}},  {\em Nucl. Phys. B} {\bf 777} (2007)
  157--181, [\href{http://arxiv.org/abs/hep-ph/0701202}{{\tt hep-ph/0701202}}].

\bibitem{Bozzi:2007tea}
G.~Bozzi, B.~Fuks, and M.~Klasen, {\it {Joint resummation for slepton pair
  production at hadron colliders}},  {\em Nucl. Phys. B} {\bf 794} (2008)
  46--60, [\href{http://arxiv.org/abs/0709.3057}{{\tt arXiv:0709.3057}}].

\bibitem{Fuks:2013lya}
B.~Fuks, M.~Klasen, D.~R. Lamprea, and M.~Rothering, {\it {Revisiting slepton
  pair production at the Large Hadron Collider}},  {\em JHEP} {\bf 01} (2014)
  168, [\href{http://arxiv.org/abs/1310.2621}{{\tt arXiv:1310.2621}}].

\bibitem{Fiaschi:2018xdm}
J.~Fiaschi and M.~Klasen, {\it {Slepton pair production at the LHC in NLO+NLL
  with resummation-improved parton densities}},  {\em JHEP} {\bf 03} (2018)
  094, [\href{http://arxiv.org/abs/1801.10357}{{\tt arXiv:1801.10357}}].

\bibitem{Fiaschi2020}
J.~Fiaschi, M.~Klasen, and M.~Sunder, {\it {Slepton pair production with
  aNNLO+NNLL precision}},  {\em JHEP} {\bf 04} (2020) 049,
  [\href{http://arxiv.org/abs/1911.02419}{{\tt arXiv:1911.02419}}].

\bibitem{Debove:2008nr}
J.~Debove, B.~Fuks, and M.~Klasen, {\it {Model-independent analysis of
  gaugino-pair production in polarized and unpolarized hadron collisions}},
  {\em Phys. Rev. D} {\bf 78} (2008) 074020,
  [\href{http://arxiv.org/abs/0804.0423}{{\tt arXiv:0804.0423}}].

\bibitem{Debove:2009ia}
J.~Debove, B.~Fuks, and M.~Klasen, {\it {Transverse-momentum resummation for
  gaugino-pair production at hadron colliders}},  {\em Phys. Lett. B} {\bf 688}
  (2010) 208--211, [\href{http://arxiv.org/abs/0907.1105}{{\tt
  arXiv:0907.1105}}].

\bibitem{Debove:2010kf}
J.~Debove, B.~Fuks, and M.~Klasen, {\it {Threshold resummation for gaugino pair
  production at hadron colliders}},  {\em Nucl. Phys. B} {\bf 842} (2011)
  51--85, [\href{http://arxiv.org/abs/1005.2909}{{\tt arXiv:1005.2909}}].

\bibitem{Debove:2011xj}
J.~Debove, B.~Fuks, and M.~Klasen, {\it {Joint Resummation for Gaugino Pair
  Production at Hadron Colliders}},  {\em Nucl. Phys. B} {\bf 849} (2011)
  64--79, [\href{http://arxiv.org/abs/1102.4422}{{\tt arXiv:1102.4422}}].

\bibitem{Fuks:2012qx}
B.~Fuks, M.~Klasen, D.~R. Lamprea, and M.~Rothering, {\it {Gaugino production
  in proton-proton collisions at a center-of-mass energy of 8 TeV}},  {\em
  JHEP} {\bf 10} (2012) 081, [\href{http://arxiv.org/abs/1207.2159}{{\tt
  arXiv:1207.2159}}].

\bibitem{Fuks:2013vua}
B.~Fuks, M.~Klasen, D.~R. Lamprea, and M.~Rothering, {\it {Precision
  predictions for electroweak superpartner production at hadron colliders with
  Resummino}},  {\em Eur. Phys. J. C} {\bf 73} (2013) 2480,
  [\href{http://arxiv.org/abs/1304.0790}{{\tt arXiv:1304.0790}}].

\bibitem{1805.11322}
J.~Fiaschi and M.~Klasen, {\it {Neutralino-chargino pair production at NLO+NLL
  with resummation-improved parton density functions for LHC Run II}},  {\em
  Phys. Rev. D} {\bf 98} (2018), no.~5 055014,
  [\href{http://arxiv.org/abs/1805.11322}{{\tt arXiv:1805.11322}}].

\bibitem{Fiaschi:2020udf}
J.~Fiaschi and M.~Klasen, {\it {Higgsino and gaugino pair production at the LHC
  with aNNLO+NNLL precision}},  {\em Phys. Rev. D} {\bf 102} (2020), no.~9
  095021, [\href{http://arxiv.org/abs/2006.02294}{{\tt arXiv:2006.02294}}].

\bibitem{1604.01023}
B.~Fuks, M.~Klasen, and M.~Rothering, {\it {Soft gluon resummation for
  associated gluino-gaugino production at the LHC}},  {\em JHEP} {\bf 07}
  (2016) 053, [\href{http://arxiv.org/abs/1604.01023}{{\tt arXiv:1604.01023}}].

\bibitem{Barger:1983wc}
V.~D. Barger, R.~W. Robinett, W.-Y. Keung, and R.~J.~N. Phillips, {\it
  {Production of Gauge Fermions at Colliders}},  {\em Phys. Lett. B} {\bf 131}
  (1983) 372.

\bibitem{Dawson:1983fw}
S.~Dawson, E.~Eichten, and C.~Quigg, {\it {Search for Supersymmetric Particles
  in Hadron - Hadron Collisions}},  {\em Phys. Rev. D} {\bf 31} (1985) 1581.

\bibitem{PhysRevLett.83.3780}
W.~Beenakker, M.~Klasen, M.~Kr\"amer, T.~Plehn, M.~Spira, and P.~M. Zerwas,
  {\it Production of charginos, neutralinos, and sleptons at hadron colliders},
   {\em Phys. Rev. Lett.} {\bf 83} (Nov, 1999) 3780--3783.

\bibitem{Berger:1999mc}
E.~L. Berger, M.~Klasen, and T.~M.~P. Tait, {\it {Associated production of
  gauginos and gluinos at hadron colliders in next-to-leading order SUSY QCD}},
   {\em Phys. Lett. B} {\bf 459} (1999) 165--170,
  [\href{http://arxiv.org/abs/hep-ph/9902350}{{\tt hep-ph/9902350}}].

\bibitem{Berger:2000iu}
E.~L. Berger, M.~Klasen, and T.~M.~P. Tait, {\it {Next-to-leading order SUSY
  QCD predictions for associated production of gauginos and gluinos}},  {\em
  Phys. Rev. D} {\bf 62} (2000) 095014,
  [\href{http://arxiv.org/abs/hep-ph/0212306}{{\tt hep-ph/0212306}}]. [Erratum:
  Phys. Rev. D 67 (2003) 099901].

\bibitem{Spira:2002rd}
M.~Spira, {\it {Higgs and SUSY particle production at hadron colliders}},  in
  {\em {10th International Conference on Supersymmetry and Unification of
  Fundamental Interactions (SUSY02)}}, pp.~217--226, 11, 2002.
\newblock \href{http://arxiv.org/abs/hep-ph/0211145}{{\tt hep-ph/0211145}}.

\bibitem{Beenakker:1994an}
W.~Beenakker, R.~Hopker, M.~Spira, and P.~M. Zerwas, {\it {Squark production at
  the Tevatron}},  {\em Phys. Rev. Lett.} {\bf 74} (1995) 2905--2908,
  [\href{http://arxiv.org/abs/hep-ph/9412272}{{\tt hep-ph/9412272}}].

\bibitem{Beenakker:1995fp}
W.~Beenakker, R.~Hopker, M.~Spira, and P.~M. Zerwas, {\it {Gluino pair
  production at the Tevatron}},  {\em Z. Phys. C} {\bf 69} (1995) 163--166,
  [\href{http://arxiv.org/abs/hep-ph/9505416}{{\tt hep-ph/9505416}}].

\bibitem{Beenakker:1996ch}
W.~Beenakker, R.~Hopker, M.~Spira, and P.~M. Zerwas, {\it {Squark and gluino
  production at hadron colliders}},  {\em Nucl. Phys. B} {\bf 492} (1997)
  51--103, [\href{http://arxiv.org/abs/hep-ph/9610490}{{\tt hep-ph/9610490}}].

\bibitem{Beenakker:1997ut}
W.~Beenakker, M.~Kramer, T.~Plehn, M.~Spira, and P.~M. Zerwas, {\it {Stop
  production at hadron colliders}},  {\em Nucl. Phys. B} {\bf 515} (1998)
  3--14, [\href{http://arxiv.org/abs/hep-ph/9710451}{{\tt hep-ph/9710451}}].

\bibitem{Kulesza:2008jb}
A.~Kulesza and L.~Motyka, {\it {Threshold resummation for squark-antisquark and
  gluino-pair production at the LHC}},  {\em Phys. Rev. Lett.} {\bf 102} (2009)
  111802, [\href{http://arxiv.org/abs/0807.2405}{{\tt arXiv:0807.2405}}].

\bibitem{Beenakker:2011sf}
W.~Beenakker, S.~Brensing, M.~Kramer, A.~Kulesza, E.~Laenen, and I.~Niessen,
  {\it {NNLL resummation for squark-antisquark pair production at the LHC}},
  {\em JHEP} {\bf 01} (2012) 076, [\href{http://arxiv.org/abs/1110.2446}{{\tt
  arXiv:1110.2446}}].

\bibitem{1601.02954}
W.~Beenakker, C.~Borschensky, R.~Heger, M.~Kr\"amer, A.~Kulesza, and E.~Laenen,
  {\it {NNLL resummation for stop pair-production at the LHC}},  {\em JHEP}
  {\bf 05} (2016) 153, [\href{http://arxiv.org/abs/1601.02954}{{\tt
  arXiv:1601.02954}}].

\bibitem{Beenakker:2009ha}
W.~Beenakker, S.~Brensing, M.~Kramer, A.~Kulesza, E.~Laenen, and I.~Niessen,
  {\it {Soft-gluon resummation for squark and gluino hadroproduction}},  {\em
  JHEP} {\bf 12} (2009) 041, [\href{http://arxiv.org/abs/0909.4418}{{\tt
  arXiv:0909.4418}}].

\bibitem{Beenakker:2013mva}
W.~Beenakker, T.~Janssen, S.~Lepoeter, M.~Kr\"amer, A.~Kulesza, E.~Laenen,
  I.~Niessen, S.~Thewes, and T.~Van~Daal, {\it {Towards NNLL resummation: hard
  matching coefficients for squark and gluino hadroproduction}},  {\em JHEP}
  {\bf 10} (2013) 120, [\href{http://arxiv.org/abs/1304.6354}{{\tt
  arXiv:1304.6354}}].

\bibitem{Beenakker:2014sma}
W.~Beenakker, C.~Borschensky, M.~Kr\"amer, A.~Kulesza, E.~Laenen, V.~Theeuwes,
  and S.~Thewes, {\it {NNLL resummation for squark and gluino production at the
  LHC}},  {\em JHEP} {\bf 12} (2014) 023,
  [\href{http://arxiv.org/abs/1404.3134}{{\tt arXiv:1404.3134}}].

\bibitem{1510.00375}
W.~Beenakker, C.~Borschensky, M.~Kr\"amer, A.~Kulesza, E.~Laenen, S.~Marzani,
  and J.~Rojo, {\it {NLO+NLL squark and gluino production cross-sections with
  threshold-improved parton distributions}},  {\em Eur. Phys. J. C} {\bf 76}
  (2016), no.~2 53, [\href{http://arxiv.org/abs/1510.00375}{{\tt
  arXiv:1510.00375}}].

\bibitem{Feng:2005gj}
J.~L. Feng, S.~Su, and F.~Takayama, {\it {Lower limit on dark matter production
  at the large hadron collider}},  {\em Phys. Rev. Lett.} {\bf 96} (2006)
  151802, [\href{http://arxiv.org/abs/hep-ph/0503117}{{\tt hep-ph/0503117}}].

\bibitem{Bai:2010hh}
Y.~Bai, P.~J. Fox, and R.~Harnik, {\it {The Tevatron at the Frontier of Dark
  Matter Direct Detection}},  {\em JHEP} {\bf 12} (2010) 048,
  [\href{http://arxiv.org/abs/1005.3797}{{\tt arXiv:1005.3797}}].

\bibitem{CMS:2017abv}
{\bf CMS} Collaboration, A.~M. Sirunyan et~al., {\it {Search for supersymmetry
  in multijet events with missing transverse momentum in proton-proton
  collisions at 13 TeV}},  {\em Phys. Rev. D} {\bf 96} (2017), no.~3 032003,
  [\href{http://arxiv.org/abs/1704.07781}{{\tt arXiv:1704.07781}}].

\bibitem{1908.04722}
{\bf CMS} Collaboration, A.~M. Sirunyan et~al., {\it {Search for supersymmetry
  in proton-proton collisions at 13 TeV in final states with jets and missing
  transverse momentum}},  {\em JHEP} {\bf 10} (2019) 244,
  [\href{http://arxiv.org/abs/1908.04722}{{\tt arXiv:1908.04722}}].

\bibitem{CMS:2021cox}
{\bf CMS} Collaboration, A.~Tumasyan et~al., {\it {Search for electroweak
  production of charginos and neutralinos in proton-proton collisions at
  $\sqrt{s} = $ 13 TeV}},  \href{http://arxiv.org/abs/2106.14246}{{\tt
  arXiv:2106.14246}}.

\bibitem{2010.14293}
{\bf ATLAS} Collaboration, G.~Aad et~al., {\it {Search for squarks and gluinos
  in final states with jets and missing transverse momentum using 139 fb$^{-1}$
  of $\sqrt{s}$ =13 TeV $pp$ collision data with the ATLAS detector}},  {\em
  JHEP} {\bf 02} (2021) 143, [\href{http://arxiv.org/abs/2010.14293}{{\tt
  arXiv:2010.14293}}].

\bibitem{2101.01629}
{\bf ATLAS} Collaboration, G.~Aad et~al., {\it {Search for squarks and gluinos
  in final states with one isolated lepton, jets, and missing transverse
  momentum at $\sqrt{s}=13$~ with the ATLAS detector}},  {\em Eur. Phys. J. C}
  {\bf 81} (2021), no.~7 600, [\href{http://arxiv.org/abs/2101.01629}{{\tt
  arXiv:2101.01629}}]. [Erratum: Eur.Phys.J.C 81, 956 (2021)].

\bibitem{hep-ph/0409313}
J.~C. Collins, D.~E. Soper, and G.~F. Sterman, {\it {Factorization of Hard
  Processes in QCD}},  {\em Adv. Ser. Direct. High Energy Phys.} {\bf 5} (1989)
  1--91, [\href{http://arxiv.org/abs/hep-ph/0409313}{{\tt hep-ph/0409313}}].

\bibitem{Gunion:1984yn}
J.~F. Gunion and H.~E. Haber, {\it {Higgs Bosons in Supersymmetric Models.
  1.}},  {\em Nucl. Phys. B} {\bf 272} (1986) 1. [Erratum: Nucl.Phys.B 402,
  567--569 (1993)].

\bibitem{Bozzi:2005sy}
G.~Bozzi, B.~Fuks, and M.~Klasen, {\it {Non-diagonal and mixed squark
  production at hadron colliders}},  {\em Phys. Rev. D} {\bf 72} (2005) 035016,
  [\href{http://arxiv.org/abs/hep-ph/0507073}{{\tt hep-ph/0507073}}].

\bibitem{Bozzi:2007me}
G.~Bozzi, B.~Fuks, B.~Herrmann, and M.~Klasen, {\it {Squark and gaugino
  hadroproduction and decays in non-minimal flavour violating supersymmetry}},
  {\em Nucl. Phys. B} {\bf 787} (2007) 1--54,
  [\href{http://arxiv.org/abs/0704.1826}{{\tt arXiv:0704.1826}}].

\bibitem{Fuks:2008ab}
B.~Fuks, B.~Herrmann, and M.~Klasen, {\it {Flavour Violation in Gauge-Mediated
  Supersymmetry Breaking Models: Experimental Constraints and Phenomenology at
  the LHC}},  {\em Nucl. Phys. B} {\bf 810} (2009) 266--299,
  [\href{http://arxiv.org/abs/0808.1104}{{\tt arXiv:0808.1104}}].

\bibitem{Fuks:2011dg}
B.~Fuks, B.~Herrmann, and M.~Klasen, {\it {Phenomenology of anomaly-mediated
  supersymmetry breaking scenarios with non-minimal flavour violation}},  {\em
  Phys. Rev. D} {\bf 86} (2012) 015002,
  [\href{http://arxiv.org/abs/1112.4838}{{\tt arXiv:1112.4838}}].

\bibitem{DeCausmaecker:2015yca}
K.~De~Causmaecker, B.~Fuks, B.~Herrmann, F.~Mahmoudi, B.~O'Leary, W.~Porod,
  S.~Sekmen, and N.~Strobbe, {\it {General squark flavour mixing: constraints,
  phenomenology and benchmarks}},  {\em JHEP} {\bf 11} (2015) 125,
  [\href{http://arxiv.org/abs/1509.05414}{{\tt arXiv:1509.05414}}].

\bibitem{Chakraborty:2018rpn}
A.~Chakraborty, M.~Endo, B.~Fuks, B.~Herrmann, M.~M. Nojiri, P.~Pani, and
  G.~Polesello, {\it {Flavour-violating decays of mixed top-charm squarks at
  the LHC}},  {\em Eur. Phys. J. C} {\bf 78} (2018), no.~10 844,
  [\href{http://arxiv.org/abs/1808.07488}{{\tt arXiv:1808.07488}}].

\bibitem{Gavin:2013kga}
R.~Gavin, C.~Hangst, M.~Kr\"amer, M.~M\"uhlleitner, M.~Pellen, E.~Popenda, and
  M.~Spira, {\it {Matching Squark Pair Production at NLO with Parton Showers}},
   {\em JHEP} {\bf 10} (2013) 187, [\href{http://arxiv.org/abs/1305.4061}{{\tt
  arXiv:1305.4061}}].

\bibitem{Hollik:2012rc}
W.~Hollik, J.~M. Lindert, and D.~Pagani, {\it {NLO corrections to squark-squark
  production and decay at the LHC}},  {\em JHEP} {\bf 03} (2013) 139,
  [\href{http://arxiv.org/abs/1207.1071}{{\tt arXiv:1207.1071}}].

\bibitem{Collins:1978wz}
J.~C. Collins, F.~Wilczek, and A.~Zee, {\it {Low-Energy Manifestations of Heavy
  Particles: Application to the Neutral Current}},  {\em Phys. Rev. D} {\bf 18}
  (1978) 242.

\bibitem{Bardeen:1978yd}
W.~A. Bardeen, A.~J. Buras, D.~W. Duke, and T.~Muta, {\it {Deep Inelastic
  Scattering Beyond the Leading Order in Asymptotically Free Gauge Theories}},
  {\em Phys. Rev. D} {\bf 18} (1978) 3998.

\bibitem{Marciano:1983pj}
W.~J. Marciano, {\it {Flavor Thresholds and Lambda in the Modified Minimal
  Subtraction Prescription}},  {\em Phys. Rev. D} {\bf 29} (1984) 580.

\bibitem{hep-ph/9308222}
S.~P. Martin and M.~T. Vaughn, {\it {Regularization dependence of running
  couplings in softly broken supersymmetry}},  {\em Phys. Lett. B} {\bf 318}
  (1993) 331--337, [\href{http://arxiv.org/abs/hep-ph/9308222}{{\tt
  hep-ph/9308222}}].

\bibitem{hep-ph/9605323}
S.~Catani and M.~H. Seymour, {\it {A General algorithm for calculating jet
  cross-sections in NLO QCD}},  {\em Nucl. Phys. B} {\bf 485} (1997) 291--419,
  [\href{http://arxiv.org/abs/hep-ph/9605323}{{\tt hep-ph/9605323}}]. [Erratum:
  Nucl.Phys.B 510, 503--504 (1998)].

\bibitem{hep-ph/0201036}
S.~Catani, S.~Dittmaier, M.~H. Seymour, and Z.~Trocsanyi, {\it {The Dipole
  formalism for next-to-leading order QCD calculations with massive partons}},
  {\em Nucl. Phys. B} {\bf 627} (2002) 189--265,
  [\href{http://arxiv.org/abs/hep-ph/0201036}{{\tt hep-ph/0201036}}].

\bibitem{hep-ph/0011222}
S.~Catani, S.~Dittmaier, and Z.~Trocsanyi, {\it {One loop singular behavior of
  QCD and SUSY QCD amplitudes with massive partons}},  {\em Phys. Lett. B} {\bf
  500} (2001) 149--160, [\href{http://arxiv.org/abs/hep-ph/0011222}{{\tt
  hep-ph/0011222}}].

\bibitem{Kinoshita1962}
T.~Kinoshita, {\it {Mass singularities of Feynman amplitudes}},  {\em J. Math.
  Phys.} {\bf 3} (1962) 650--677.

\bibitem{Lee1964}
T.~D. Lee and M.~Nauenberg, {\it {Degenerate Systems and Mass Singularities}},
  {\em Phys. Rev.} {\bf 133} (1964) B1549--B1562.

\bibitem{hep-ph/0010146}
A.~Vogt, {\it {Next-to-next-to-leading logarithmic threshold resummation for
  deep inelastic scattering and the Drell-Yan process}},  {\em Phys. Lett. B}
  {\bf 497} (2001) 228--234, [\href{http://arxiv.org/abs/hep-ph/0010146}{{\tt
  hep-ph/0010146}}].

\bibitem{Kramer:1996iq}
M.~Kramer, E.~Laenen, and M.~Spira, {\it {Soft gluon radiation in Higgs boson
  production at the LHC}},  {\em Nucl. Phys. B} {\bf 511} (1998) 523--549,
  [\href{http://arxiv.org/abs/hep-ph/9611272}{{\tt hep-ph/9611272}}].

\bibitem{Harlander:2001is}
R.~V. Harlander and W.~B. Kilgore, {\it {Soft and virtual corrections to proton
  proton ---\ensuremath{>} H + x at NNLO}},  {\em Phys. Rev. D} {\bf 64} (2001)
  013015, [\href{http://arxiv.org/abs/hep-ph/0102241}{{\tt hep-ph/0102241}}].

\bibitem{Catani:2001ic}
S.~Catani, D.~de~Florian, and M.~Grazzini, {\it {Higgs production in hadron
  collisions: Soft and virtual QCD corrections at NNLO}},  {\em JHEP} {\bf 05}
  (2001) 025, [\href{http://arxiv.org/abs/hep-ph/0102227}{{\tt
  hep-ph/0102227}}].

\bibitem{Kulesza:2002rh}
A.~Kulesza, G.~F. Sterman, and W.~Vogelsang, {\it {Joint resummation in
  electroweak boson production}},  {\em Phys. Rev. D} {\bf 66} (2002) 014011,
  [\href{http://arxiv.org/abs/hep-ph/0202251}{{\tt hep-ph/0202251}}].

\bibitem{Almeida:2009jt}
L.~G. Almeida, G.~F. Sterman, and W.~Vogelsang, {\it {Threshold Resummation for
  Di-hadron Production in Hadronic Collisions}},  {\em Phys. Rev. D} {\bf 80}
  (2009) 074016, [\href{http://arxiv.org/abs/0907.1234}{{\tt
  arXiv:0907.1234}}].

\bibitem{Contopanagos:1993yq}
H.~Contopanagos and G.~F. Sterman, {\it {Principal value resummation}},  {\em
  Nucl. Phys. B} {\bf 419} (1994) 77--104,
  [\href{http://arxiv.org/abs/hep-ph/9310313}{{\tt hep-ph/9310313}}].

\bibitem{Catani:1996yz}
S.~Catani, M.~L. Mangano, P.~Nason, and L.~Trentadue, {\it {The Resummation of
  soft gluons in hadronic collisions}},  {\em Nucl. Phys. B} {\bf 478} (1996)
  273--310, [\href{http://arxiv.org/abs/hep-ph/9604351}{{\tt hep-ph/9604351}}].

\bibitem{ParticleDataGroup:2020ssz}
{\bf Particle Data Group} Collaboration, P.~A. Zyla et~al., {\it {Review of
  Particle Physics}},  {\em PTEP} {\bf 2020} (2020), no.~8 083C01.

\bibitem{lhapdf}
A.~Buckley, J.~Ferrando, S.~Lloyd, K.~Nordstr\"om, B.~Page, M.~R\"ufenacht,
  M.~Sch\"onherr, and G.~Watt, {\it {LHAPDF6: parton density access in the LHC
  precision era}},  {\em Eur. Phys. J. C} {\bf 75} (2015) 132,
  [\href{http://arxiv.org/abs/1412.7420}{{\tt arXiv:1412.7420}}].

\bibitem{Bailey:2020ooq}
S.~Bailey, T.~Cridge, L.~A. Harland-Lang, A.~D. Martin, and R.~S. Thorne, {\it
  {Parton distributions from LHC, HERA, Tevatron and fixed target data: MSHT20
  PDFs}},  {\em Eur. Phys. J. C} {\bf 81} (2021), no.~4 341,
  [\href{http://arxiv.org/abs/2012.04684}{{\tt arXiv:2012.04684}}].

\bibitem{1405.0301}
J.~Alwall, R.~Frederix, S.~Frixione, V.~Hirschi, F.~Maltoni, O.~Mattelaer,
  H.~S. Shao, T.~Stelzer, P.~Torrielli, and M.~Zaro, {\it {The automated
  computation of tree-level and next-to-leading order differential cross
  sections, and their matching to parton shower simulations}},  {\em JHEP} {\bf
  07} (2014) 079, [\href{http://arxiv.org/abs/1405.0301}{{\tt
  arXiv:1405.0301}}].

\bibitem{hep-ph/0202233}
B.~C. Allanach et~al., {\it {The Snowmass Points and Slopes: Benchmarks for
  SUSY Searches}},  {\em Eur. Phys. J. C} {\bf 25} (2002) 113--123,
  [\href{http://arxiv.org/abs/hep-ph/0202233}{{\tt hep-ph/0202233}}].

\bibitem{Porod:2011nf}
W.~Porod and F.~Staub, {\it {SPheno 3.1: Extensions including flavour,
  CP-phases and models beyond the MSSM}},  {\em Comput. Phys. Commun.} {\bf
  183} (2012) 2458--2469, [\href{http://arxiv.org/abs/1104.1573}{{\tt
  arXiv:1104.1573}}].

\bibitem{Pumplin:2002vw}
J.~Pumplin, D.~R. Stump, J.~Huston, H.~L. Lai, P.~M. Nadolsky, and W.~K. Tung,
  {\it {New generation of parton distributions with uncertainties from global
  QCD analysis}},  {\em JHEP} {\bf 07} (2002) 012,
  [\href{http://arxiv.org/abs/hep-ph/0201195}{{\tt hep-ph/0201195}}].

\bibitem{1710.11091}
E.~Bagnaschi et~al., {\it {Likelihood Analysis of the pMSSM11 in Light of LHC
  13-TeV Data}},  {\em Eur. Phys. J. C} {\bf 78} (2018), no.~3 256,
  [\href{http://arxiv.org/abs/1710.11091}{{\tt arXiv:1710.11091}}].

\bibitem{Muong-2:2021ojo}
{\bf Muon g-2} Collaboration, B.~Abi et~al., {\it {Measurement of the Positive
  Muon Anomalous Magnetic Moment to 0.46 ppm}},  {\em Phys. Rev. Lett.} {\bf
  126} (2021), no.~14 141801, [\href{http://arxiv.org/abs/2104.03281}{{\tt
  arXiv:2104.03281}}].

\bibitem{pdfmonte}
J.~Butterworth et~al., {\it {PDF4LHC recommendations for LHC Run II}},  {\em J.
  Phys. G} {\bf 43} (2016) 023001, [\href{http://arxiv.org/abs/1510.03865}{{\tt
  arXiv:1510.03865}}].

\bibitem{Hou:2019efy}
T.-J. Hou et~al., {\it {New CTEQ global analysis of quantum chromodynamics with
  high-precision data from the LHC}},  {\em Phys. Rev. D} {\bf 103} (2021),
  no.~1 014013, [\href{http://arxiv.org/abs/1912.10053}{{\tt
  arXiv:1912.10053}}].

\bibitem{Ball:2021leu}
R.~D. Ball et~al., {\it {The Path to Proton Structure at One-Percent
  Accuracy}},  \href{http://arxiv.org/abs/2109.02653}{{\tt arXiv:2109.02653}}.

\bibitem{2104.09174}
J.~Rojo, {\it {Progress in the NNPDF global analyses of proton structure}},  in
  {\em {55th Rencontres de Moriond on QCD and High Energy Interactions}}, 4,
  2021.
\newblock \href{http://arxiv.org/abs/2104.09174}{{\tt arXiv:2104.09174}}.

\bibitem{resumminourl}
``Resummino.'' \url{https://resummino.hepforge.org/}.

\bibitem{Fuks:2007gk}
B.~Fuks, M.~Klasen, F.~Ledroit, Q.~Li, and J.~Morel, {\it {Precision
  predictions for $Z^\prime$ - production at the CERN LHC: QCD matrix elements,
  parton showers, and joint resummation}},  {\em Nucl. Phys. B} {\bf 797}
  (2008) 322--339, [\href{http://arxiv.org/abs/0711.0749}{{\tt
  arXiv:0711.0749}}].

\bibitem{rothering:phd}
M.~Rothering, {\em {Precise predictions for supersymmetric particle production
  at the LHC }}.
\newblock dissertation, Westfälischen Wilhelms-Universität Münster, 2016.

\bibitem{theeuwe:phd}
V.~M. Theeuwes, {\em {Soft Gluon Resummation for Heavy Particle Production at
  the Large Hadron Collider}}.
\newblock dissertation, Westfälischen Wilhelms-Universität Münster, 2015.

\bibitem{Kidonakis:2006bu}
N.~Kidonakis, {\it {Single top production at the Tevatron: Threshold
  resummation and finite-order soft gluon corrections}},  {\em Phys. Rev. D}
  {\bf 74} (2006) 114012, [\href{http://arxiv.org/abs/hep-ph/0609287}{{\tt
  hep-ph/0609287}}].

\bibitem{Laenen:1998qw}
E.~Laenen, G.~Oderda, and G.~F. Sterman, {\it {Resummation of threshold
  corrections for single particle inclusive cross-sections}},  {\em Phys. Lett.
  B} {\bf 438} (1998) 173--183,
  [\href{http://arxiv.org/abs/hep-ph/9806467}{{\tt hep-ph/9806467}}].

\end{thebibliography}\endgroup

\end{document}